\documentclass[11pt]{article}

\usepackage{amsmath}
\usepackage{amssymb}
\usepackage{epsfig}

\newtheorem{theorem}{Theorem}[section]
\newtheorem{proposition}{Proposition}[section]

\begin{document}


\begin{titlepage}

\renewcommand{\thefootnote}{\alph{footnote}}
\vspace*{-3.cm}
\begin{flushright}

\end{flushright}

\vspace*{0.5cm}

\renewcommand{\thefootnote}{\fnsymbol{footnote}}
\setcounter{footnote}{-1}

{\begin{center} {\Large\bf The  Euler-Riemann Gases, and Partition
Identities}

\end{center}}
\renewcommand{\thefootnote}{\alph{footnote}}

\vspace*{.8cm} {\begin{center} {\large{\sc
                Noureddine~Chair
                }}
\end{center}}
\vspace*{0cm} {\it
\begin{center}
 \begin{center}
 Physics Department,
 The University of Jordan, Amman, Jordan 
 \begin{center}

     Email: n.chair@ju.edu.jo\\ \hspace{19mm}\\
\end{center}
\end{center}
\end{center} }

\vspace*{1.5cm}

{\Large \bf
\begin{center} Abstract \end{center} }
The Euler theorem  in partition theory and its generalization are derived from a non-interacting  quantum field theory in which each bosonic mode with a given frequency is equivalent to a sum of bosonic mode whose frequency is twice ($s$-times) as much, and a fermionic (parafermionic) mode with the same frequency.   Explicit formulas for the graded
parafermionic partition functions are obtained, and  the inverse of the graded partition function (IGPPF), turns out to be  bosonic (fermionic)
partition function depending on the parity of the order $s$ of the parafermions. It is also shown that these partition functions are
generating functions of partitions of integers with restrictions,
the Euler generating function is identified with the inverse of the 
graded parafermionic partition function of order 2. As a result we
obtain new sequences of partitions of integers with given
restrictions. If the parity of the order $s$ is even, then  mixing a system of parafermions 
with a system whose partition function is (IGPPF), results in a system of fermions
and bosons.  On the other hand, if the parity of $s$ is odd, then, the system we obtain is still
a mixture of fermions and bosons but the corresponding Fock space
of states is truncated. It turns out that these partition
functions are given in terms of the Jacobi theta function
$\theta_{4}$, and generate sequences in partition theory.  Our
partition functions coincide with the overpartitions of Corteel
and Lovejoy, and jagged partitions in conformal field theory. Also, The partition functions  obtained are related to the Ramond characters of the superconformal  minimal models, and in the counting of  the Moore-Read edge spectra that appear in the fractional quantum Hall effect. The different partition functions for the Riemann gas that are the counter parts of the Euler gas are obtained  by a simple change of variables. In particular the counter part of the Jacobi Theta
function is $\frac{\zeta(2t)}{\zeta(t)^2}$. Finally, We propose two formulas which brings  the additive number theory and the multiplicative number theory closer.

\vspace*{.5cm}

\end{titlepage}

\newpage

\renewcommand{\thefootnote}{\arabic{footnote}}
\setcounter{footnote}{0}


%

\section{INTRODUCTION \label{sec:SEC-intro}}
In arithmetic quantum theories the spectrum is chosen to be
logarithmic in order to connect quantum mechanics to number
theory (multiplicative number theory), since the partition
function is related to the Riemann zeta function or other Dirichlet series
\cite{Julia} and to string theory \cite{i.bakas}. In these
theories, number theoretic identities have been derived and
interpreted. Spector in \cite{Spector} obtained the expressions for the fermionic and parafermionic partition function in terms of a bosonic partition function using the notion of
partial supersymmetry. Although his expressions for the partition functions are written correctly, his assumption that the powers  of  bosonic operators are  bosonic operators turns out to be false. This problem is fixed by saying that the spectrum  of a single bosonic mode (harmonic oscillator) is a sum of a spectrum of an even harmonic oscillator and that of a fermionic oscillator.  Keeping  this interpretation in mind, a non-interacting quantum theory free of a logarithmic spectrum is then considered and  shown to be  connected  to the Euler generating function for
partitions. As  a consequence,  the quantum theory we considered is connected with the additive number theory \cite{apostol}.  In this paper, the Riemann gas is also considered by relating the different partition  functions of Euler gas additively computed with those of the Riemann gas that  has logarithmic spectrum. We have derived the Euler theorem by expressing the 
fermionic partition function in terms of the bosonic partition function, the Euler theorem says that the number of
partitions of a number $k$ (a positive integer) containing odd numbers only equals the
number of partitions of a number $k$ without duplication. It turns out that the square of a bosonic operator is related  to a parafermion  whose order is not an integer\cite{Kademova}. The commutation relations satisfied by these operators is a modification of the Green's  trilinear relations \cite{Green}. Therefore, the quantum theory that we should consider is constructed in such a way  that it does not contain the square of the bosonic operators. The bosonic  partition function of a single bosonic  oscillator in this case, is the sum of two parafermionic partition functions, one with parastatistics order $s=1/2$ and the other with order $s=3/2$ . This sum then tuns into the  product of a bosonic and fermionic partition functions at different temperatures. 
Similarly, from parafermionic partition function we have derived
another identity in partition theory \cite{andrews} which equates
the number of partitions of $k$ in which no part appears more than
$s-1$ times, with the number of partitions of $k$ such that no
part is divisible by $s$. Using this identity we prove Andrews'
result \cite{andrews} in connection with generating functions that
exclude squares and their generalizations \cite{sellers}. By considering the
graded parafermionic case, some explicit formulas  for
the partition functions are obtained. It turns out that the partition functions that have positive norms are  the inverse of the graded parafermionic partition functions  (IGPPF). These partition functions  correspond to the
generalized Euler generating function for partitions and give rise to  sequences of
partitions of integers that have appeared in \cite{sloane}. The Euler's generating
function of partitions  has the interpretation of being the
inverse of the  graded fermionic partition function. This is the
partition function of the Euler gas, it will be shown that its
generalization  is a bosonic (fermionic) partition function
depending on the parity of the order $s$  of parafermions. Mixing a
parafermionic system with a system whose partition function is  (IGPPF) results in mixing
fermions with bosons. The corresponding Fock space of the states
associated to this mixture is a tensor product (convolution) of the
fermoinic and the bosonic Fock spaces. This happens when the order
of parafermions is even, however, if the order is odd, then the
mixing is still made of fermions and bosons but the corresponding
Fock space of states is truncated. It turns out that these
partition functions are related to the Jacobi theta function and
have a meaning in partition theory and are identical to the theory  of  overpartitions
\cite{corteel}, or jagged partitions  in conformal field theory \cite{Fortin1} and their restrictions. These partition functions  are related to the Ramond characters of the superconformal  minimal models \cite{Fortin2}, and in the counting of  the Moore-Read edge spectra that appear in the fractional quantum Hall effect \cite{Michael}.  The different
partition functions obtained in this paper are related to
generating functions of partitions of integers. therefore,  we may
call these theories additive quantum theories to differentiate
them from multiplicative quantum theories whose partition
functions are related to the Riemann zeta function. By using the fact that
bosonic partition functions for the Riemann gas and the Euler gas
are given by products. Then, it is possible to go from the additive quantum
theory to multiplicative quantum theory, and write down all the different partition functions for the Riemann gas by a simple change of variables.  All these partition functions are
written in terms of the Riemann zeta functions, or alternatively,  written  as Dirichlet series. Our paper is organized as follows, in sections 2 and 3, the Euler theorem in partition theory and its generalization are proved by using a  non-interacting quantum field theory. Here, in this theory, each bosonic mode of a given  frequency is considered as a sum of a bosonic mode whose frequency is twice ($s$-times) as much  and fermionic (parafermionic) mode with the same frequency. In Section 4, the graded parafermionc  partition function, i.e, generalization of the graded fermionic partition function, and  the inverse of the graded parafermionic partition functions are considered. These partition functions are  bosonic (fermionic) partition functions  depending on the parity of the order $ s$ of the parafermion. Here, we gave a number of examples of generating function of sequences of partitions, some of which are the outcome of the present work. In Section 5,  partition functions  of a mixed system made of parafermions system and  bosonic (fermionic ) system are shown to be related the Jacobi Theta function, $ \theta_{4}(0,x)$. The partition functions of the Riemann gas and their relation to those of the Euler gas are   discussed in great details in Section 6, in particular, it is shown that the counter part of the partition function $\frac{1}{ \theta_{4}(0,x)} $ in the additive quantum theory, is $\frac{\zeta(t)^2}{\zeta(2t)}$. $\theta_{4}$ is  like the theta function $\theta $ is in the ordinary variables  additive in nature, while the Dirichlet and in particular the Riemann zeta function are multiplicative. The Mellin transform  is used  to translate additive into multiplicative and vice versa. It is well known that the Riemann zeta function is the Mellin transform of $(\theta(0,x) -1)$. A simple exercise shows that the alternating Riemann zeta function or the Dirichlet eta function is the Mellin transform of $\theta_{4}$. Finally, in Section 7,  a detailed discussion is given in which  we propose a number of formulas that makes the additive and multiplicative number theories have certain resemblance. Also, some open problems are given  related to the  the parafermionic partition functions.
\section{ PARTITION FUNCTIONS AND THE EULER IDENTITY}
Here, we  derive the Euler identity from the  partition function of a bosonic non-interacting quantum field theory. The  Euler identity says; the generating functions for the number of partitions of a given number $k$ (a positive integer) into distinct parts and the number of partitions of $k$ into odd parts are equal. Our derivation is based on the construction of the bosonic Hamiltonian of a non-interacting quantum field theory in terms of some bosonic operators and fermionic operators. This construction is close to that given by Spector \cite{Spector} on partial supersymmetry.  Spector \cite{Spector}, he assumed that   the bosonic operators were given by $q_k=(b_k)^2$ and $q_k^{\dag}=(b_k^{\dag})^2$, where $b_k^{\dag}(b_k)$, are the  bosonic creation(annihilation) operators respectively. However, a simple computation shows that the commutator of these operators is  not a $c$-number, $[ (b_{k})^{2}_ ,( b_{m}^{\dag})^{2}]=\delta_{km}(4b_k^{\dag}b_k+2)$. Therefore, the operators $q_k=(b_k)^2$ and $q_k^{\dag}=(b_k^{\dag})^2$ are not bosonic operators. In order to motivate our decomposition of the bosonic Hamiltonian, we note that for  a single bosonic  mode (harmonic oscillator), the Hamiltonian is  $H=\omega b^{\dag}b$, $\omega $ beeing the frequency of the oscillator, then the energy may be written as  $E_{n}=n\omega =2n\omega+(2n+1)\omega $, $n=0, 1,2,\cdots$. Thus, the spectrum of $H$ splits into even and odd harmonics. This simple expression says that the energy $E_{n}$ may be obtained by adding to the energy of an even bosonic harmonic  oscillator, the energy of a fermionic oscillator that has  a zero frequency and a single frequency $\omega$ in its spectrum. As a consequence, the Hamiltonian of a single bosonic mode may decomposed as $H= 2\omega q^{\dag}q +\omega c^{\dag}c $, where  $q^{\dag}(q)$, are the bosonic creation(annihilation) operators for the even harmonic oscillator  and $c^{\dag}(c)$, are the creation (annihilation) operators that have the same effect as $b$ and
$b^{\dag}$ respectively, the difference being that they are
square free (fermionic operators), that is, $(c)^{2}=( c^{\dag})^2 =0$. This decomposition may extended  to all the bosonic modes in the non-interacting  quantum field theory. The next four sections in this paper are connected  to non-interacting quantum field
theories whose spectrum is not logarithmic, and so we can
write the bosonic (fermionic) Hamiltonian without the logarithm of
a prime in the form,
\begin{equation}
\label{toto1}
H_B=\omega\sum_{k=1}^\infty k b_k^{\dag}b_k.
\end{equation}
\begin{equation}
\label{toto2}
H_F=\omega\sum_{k=1}^\infty k f_k^{\dag}f_k
\end{equation}
where $b_k^{\dag}(b_k)$, are the bosonic creation(annihilation)
operators respectively and $f_k^{\dag}(f_k)$, are the fermionic
creation (annihilation) operators respectively.the $k^{th}$ harmonic oscillator has a frequency $k\omega$. The operators $b_k$ and $b_k^{\dag}$
satisfy  the usual commutators given by $[ b_m , b_n^{\dag} ]
=\delta_{nm}$. The ground state $|\emptyset\rangle$ is defined by
$b_k |\emptyset\rangle=0$, for all $k$ . The quantum states 
denoted by $|n\rangle$ are obtained by acting with the
creation operators on the ground state,i.e,
$|n\rangle=(b_1^{\dag})^{n_1}(b_2^{\dag})^{n_2}\cdots
(b_k^{\dag})^{n_k}\cdots|\emptyset\rangle$. Applying the number
operator $\textbf{n}=\sum_{k=1}^\infty k b_k^{\dag}b_k$ on the state
$|n\rangle$, we have $\textbf{n}|n\rangle=n |n\rangle$ where
$n=\sum_{k=1}^\infty kn_k$. For fermions, the commutators are given by $\{ f_k
, f_m^{\dag} \} =\delta_{km}$, in this case  the allowed eigenvalues of the number operator $f_k^{\dag}f_k$ are $n_k=1$ or $0$. Just like we did for a single mode, the bosonic Hamiltonian may be decomposed as
\begin{equation}
\label{toto3}
H_B=\omega\sum_{k=1}^\infty
kc_k^{\dag}c_k+\omega\sum_{k=1}^\infty 2kq_k^{\dag}q_k,
\end{equation}
where  $q_k$,$q_k^{\dag}$ are the bosonic creation (annihilation) operators in the even bosonic oscillators,  $(c_k)^2=(c_k^{\dag})^2=0$ are the square free operators which have the same effect as $q_k$,$q_k^{\dag}$.  As a result the $c_{k}$ partition function is not graded, that is, $$ {\rm
Tr}(-1)^F\exp(-\beta\omega\sum_{k=1}^\infty kc_k^{\dag}c_{k})={\rm
Tr}\exp(-\beta\omega\sum_{k=1}^\infty kc_k^{\dag}c_{k}),$$  where $F$ is the fermion number operator  with eigenvalues 0 or 1. It is equal to $0$ on the  states created by bosonic operators. The fermionic partition function may be obtained using the following identity 
\begin{eqnarray}{\rm
Tr}\left[(-1)^F\exp{-\beta(H_B+H_F)}\right]={\rm
Tr}\exp{-\beta H_B}{\rm
Tr}(-1)^F\exp {-\beta H_F}=1.\nonumber
\end{eqnarray}
This identity may be checked using the expressions for $H_{B}$ and $H_{F}$, formally, this is equivalent to the  Witten index \cite{witten}. 
Now, since the components of our  Hamiltonian
$H_B$  commute, then,  the trace in the bosonic partition function
decomposes as
\begin{equation}
\label{toto4}
{\rm Tr}\left[\exp{-\beta H_B}\right]= {\rm
Tr}\left[\exp{-\beta\omega\sum_{k=1}^\infty kc_k^{\dag}c_k}\right]
{\rm Tr}\left[\exp{-\beta\omega\sum_{k=1}^\infty
2kq_k^{\dag}q_k}\right],
\end{equation}
where the first term on the right of this equation is the fermionic
partition function and the second term is the bosonic partition
function. Adding the term $\omega\sum_{k=1}^\infty
2kf_k^{\dag}f_k$ to the original Hamiltonian and then using the identity  equivalent to the 
Witten index, then the  fermionic  partition may be written as 
\begin{eqnarray}
\label{toto5}
Z_F(\beta)&=&{\rm
Tr}\left[\exp(-\beta\omega\sum_{k=1}^\infty kc_k^{\dag}c_{k})\right]=
{\rm Tr}\left[(-1)^F\exp{-\beta\omega\sum_{k=1}^\infty
(kc_k^{\dag}c_k+2kq_k^{\dag}q_k+2kf_k^{\dag}f_k)}\right]\nonumber\\
&=& {\rm Tr}\left[(-1)^F\exp{-\beta(H_B+2H_F)}\right].
\end{eqnarray}
The bosonic and the fermionic partition functions in equation
(\ref{toto5}) are computed in the Fock space of states for both
bosons and fermions using the expressions for $H_B$ and $H_F$
given by equation (\ref{toto1}) and equation (\ref{toto2})
respectively, and so we have,
\begin{eqnarray}
\label{toto6}
Z_B(\beta)&=& {\rm Tr}\left[\exp{-\beta
H_B}\right]={\rm Tr}\left[\exp{-\beta\omega\sum_{k=1}^\infty
kb_k^{\dag}b_k}\right]=\prod_{k=1}^\infty\sum_{n_{k}=0}^\infty x^
{n_{k}k}=\prod_{k=1}^{\infty}\frac{1}{1-x^k},
\end{eqnarray}
where we have set $x=\exp{-(\beta\omega)}$, one can easily check
that the partition function for equation (\ref{toto3}) and that
for $H_{B}$ equation (\ref{toto1}) agree. The computation in the
fermionic case is similar except that due to the Fermi-Dirac
statistics $n_{k}$ takes the values $0$ and $1$, therefore,
\begin{eqnarray}
\label{toto7}
Z_F(\beta)&=&{\rm
Tr}\left[\exp{-\beta\omega\sum_{k=1}^\infty
kc_k^{\dag}c_k}\right]=\prod_{k=1}^\infty\sum_{n_{k}=0}^1 x^
{n_{k}k}=\prod_{k=1}^{\infty}({1+x^k}).
\end{eqnarray}
The remaining partition function to be computed is the graded
fermionic partition \begin{eqnarray} {\rm
Tr}\left[(-1)^F\exp{-\beta(2H_F)}\right] \nonumber.
\end{eqnarray}
Since the fermion number (the eigenvalue of $F$) is either $0$
or $1$, then if we denote this partition by $\Delta_F(\beta)$ we
have
\begin{eqnarray}
\label{toto8}
\Delta_F(2\beta)&=&{\rm
Tr}\left[(-1)^F\exp(-2\beta\omega\sum_{k=1}^{\infty}
kf_{k}^{\dag}f_{k})\right]=\prod_{k=1}^{\infty}\sum_{n_{k}=0}^{1}(-1)^{n_{k}}
x^{2n_{k}k}=\prod_{k=1}^{\infty}(1-x^{2k}).
\end{eqnarray}
Finally, the factorization identity gives

\begin{eqnarray}
\label{toto9}
\prod_{k=1}^{\infty}({1+x^k})&=&
\prod_{k=1}^{\infty}\frac{(1-x^{2k})}{({1-x^k})}=
\prod_{k=1}^{\infty}\frac{(1-x^{2k})}{(1-x^{2k})(1-x^{2k-1})}=
\prod_{k=1}^{\infty}\frac{1}{(1-x^{2k-1})}.
\end{eqnarray}
This is a well known theorem due to Euler \cite{apostol} which
says; that the generating functions for the number of partitions of
a given number $k$ into distinct parts and the number of
partitions of $k$ into odd parts are equal. Here, we have obtained
this equality from a fermionic partition function that is part of the full bosonic partition function.  Next, we will write down
the partition functions for the bosonic and fermionic simple
harmonic oscillators respectively. These two partition functions are related to each other through the identity  given by equation (\ref{toto5}), and  as a result we get a well known identity
for $\cosh{x}$ in terms of an infinite  product. The Hamiltonian
for the bosonic (fermionic) simple harmonic oscillator are
$H_B=\omega (b{\dag}b+{1/2}) $( $H_F=\omega (f^{\dag}f-{1/2}))$
respectively and hence their partition functions are,
\begin{eqnarray}
\label{toto10}
Z_B(\beta) &=&{\rm
Tr}\left[\exp(-{\beta\omega}(b^{\dag}b+{1/2}))\right] \nonumber\\
&=& \sum_{n=0}^{\infty}\exp(-\beta\omega(n+{1/2}))=
\frac{1}{2\sinh(\beta\omega/2)}.
\\\label{toto11}
Z_F(\beta) &=&{\rm
Tr}\left[\exp(-\beta\omega(f^{\dag}f -{1/2}))\right]\nonumber\\
&=&\sum_{n=0}^{1}\exp\left[- \beta\omega(n-  {1/2})\right] =
2(\cosh(\beta\omega/2)),
\end{eqnarray}
The other piece we need to compute is the graded fermionic partition
function $$\Delta_{F}(2\beta)= {\rm
Tr}\left[(-1)^{F}\exp(-\beta\omega(f^{\dag}f-{1/2}))\right]$$ which
is simple to calculate since the eigenvalues of the fermion number
operator $F$ are $0$ and $1$ and so we have
$\Delta_{F}(2\beta)=2\sinh(\beta\omega/2)$ since the infinite
product representation of  sine hyperbolic is
$\sinh(x)=x\prod_{k=0}^{\infty}(1+\frac{x^{2}}{k^{2}\pi^{2}})$. Then
the factorization identity (\ref{toto5}) gives
\begin{eqnarray}
\label{toto12}
Z_F(\beta)&=&2\cosh(\beta\omega)/2)=\frac{2\sinh(\beta\omega)}{2\sinh(\beta\omega/2)}\nonumber\\&=&
2\prod_{k=1}^{\infty}\frac{(1+\frac{(\beta\omega)^{2}}{k^{2}\pi^{2}})}{(1+\frac{(\beta\omega)^{2}}{4(k)^{2}\pi^{2}})}=
2\prod_{k=1}^{\infty}{(1+\frac{(\beta\omega)^{2}}{(2k+1)^{2}\pi^{2}})},
\end{eqnarray}
which is exactly the infinite product representation of
$(\cosh(\beta\omega)/2)$, therefore, it follows that  the results one
obtains by writing the fermionic partition function in terms of a bosonic partition function depends very much on the
Hamiltonian used. If one is dealing with a theory with a
logarithmic spectrum, then this procedure  gives a proof of a
number theoretic identity that connects the zeta function to the
modulus of Mobius inversion function \cite{Spector}. Here, we gave the generating function proof of the Euler
theorem in which the number of partitions of $k$ into distinct
parts equals the number of partitions of $k$ into odd parts. Also,
in the case the Hamiltonian is that of a harmonic oscillator, we
have derived a well known  identity in hyperbolic trigonometry
$2\cosh(\beta\omega/2)=\frac{\sinh(\beta\omega)}{\sinh(\beta\omega/2)}$,
from which we obtain the infinite product representation of
$\cosh{x}$ knowing that of $\sinh(x)$
\section{PARAFERMIONIC PARTITION FUNCTIONS AND PARTITIONS WITH RESTRICTIONS}
The natural generalization to the previous section, in which the
fermionic partition function was factorized as a product of graded
fermionic partition function $\Delta_F(2\beta)$ times the bosonic
partition function $Z_B(\beta)$, would be to consider parafermions
of order s. As fermions are of order $2$ parafermions, therefore, just like the
first factorization identity (\ref{toto5}),  we will have the
following second factorization identity
\begin{eqnarray}
\label{toto13} Z_s(\beta)&=& {\rm
Tr}\left[(-1)^F\exp{-\beta(H_B+sH_F)}\right],\nonumber\\
&=&Z_B(\beta)\Delta_F(s\beta)
\end{eqnarray}
where the Hamiltonian $H_B$, is constructed out of certain operators
$\chi_k$ and $r_k$ such that $(\chi_k)^s=(\chi_k{\dag})^s=0$ but no
lower powers vanish as operators, i.e., these are parafermionic
operators, the  operators  $ r_k$,
$r_k^{\dag}$ are the operators of  the bosonic oscillators  having frequencies with multiplicity $s$. These oscillators are the natural generalization of the even bosonic oscillators considered in the last section. Spector \cite{Spector}, in his construction of the bosonic Hamiltonian, he assumed that the  operators  given by  $ r_k=(b_{k})^s$, $r_k^{\dag}=(b_{k}^{\dag})^s$ are bosonic operators. However, from the following explicite expressions of the  commutators for $s=3$, $s=4$, $$[ (b)^{3} , (b^{\dag})^3]=9b^{\dag}bb^{\dag}b+9b^{\dag}b+6,$$ $$[ (b)^{4},( b^{\dag})^{4}]=16b^{\dag}bb^{\dag}bb^{\dag}b+24b^{\dag}bb^{\dag}b+56b^{\dag}b+24,$$ it is clear  that the powers of the bosonic operators are not bosonic operators. Now, let us see how to construct the  Hamiltonian of a bosonic oscillator in terms of a parafermion of order three. The energy of the  bosonic oscilator in this case may be written as $E_{n}= \omega n=\omega(3n)+\omega(3n+1)+\omega(3n+2)$, $n=0,1,2,\cdots$, since both sides of this expresion give the same spectrum  of the Hamiltonian $H_{B}=\omega b^{\dag}b$.  As a consequence, the Hamiltonian of a single bosonic mode may decomposed as $H_{B}= 3\omega r^{\dag}r +\omega \chi^{\dag}\chi $, where  $r^{\dag}(r)$, are the bosonic creation (annihilation) operators for the  harmonic oscillator whose frequency is $3\omega$, and $\chi^{\dag}(\chi)$, are the creation (annihilation) operators that have the same effect as $b^{\dag}(b)$ respectively, except that $(\chi^{\dag})^{3}=(\chi)^{3}=0$, that is, cubic free. Note that the eigenvalues of the number operator $\chi^{\dag}\chi$  for $s=3$ are $n=0, 1, 2 $, and for general $s$, the eigenvalues are  $n=0, 1,2,\cdots, s-1$. In general, the decomposition of the bosonic Hamiltonian in the non-interacting  quantum field theory may be witten as  
\begin{equation}
\label{toto14}
 H_B=\omega\sum_{k=1}^\infty \chi_k^{\dag}\chi_k+\omega\sum_{k=1}^\infty sr_k^{\dag}r_k.
\end{equation}
 By the using the identity $$ {\rm
Tr}\left[(-1)^F\exp{-\beta(\omega\sum_{k=1}^\infty sr_k^{\dag}r_k+s\omega\sum_{k=1}^\infty k f_k^{\dag}f_k)}\right]=1,$$ which is equivalent to the Witten index, then the  parafermionic partition reads,
\begin{equation}
\label{toto15} Z_s(\beta)= {\rm Tr}\exp(-\beta(H_{s}))={\rm
Tr}\left[(-1)^{F}\exp(-\beta(H_B+sH_F))\right]
\end{equation}
where $H_{s}=\omega\sum_{k=1}^{\infty} \chi_{k}^{\dag}\chi_k$. The
parafermionic partition function is a sort of truncated bosonic
partition since the term $x^{sk}$ and higher terms are not present
and so $Z_{s}(\beta)=  \prod_{k=1}^{\infty}(1+ x^{k} + x^{2k}+
\cdots + x^{(s-1)k})$ and as a result by using the second
factorization identity, we obtain

\begin{eqnarray}
\label{toto16}
Z_{s}(\beta)&=&{\rm Tr}\exp(-\beta(H_{s}))=\sum_{n_{k}=0}^{s-1} x^
{\sum_{k=1}^{\infty} k n_{k}}\nonumber\\&=&\prod_{k=1}^{\infty}(1+ x^{k} + x^{2k}+
\cdots +
x^{(s-1)k})\nonumber\\
&=&\prod_{k=1}^{\infty}\frac{(1-x^{sk})}{({1-x^k})}.
\end{eqnarray}
This is exactly an identity in the theory of partitions
\cite{andrews} which says that the the generating function for the
number of partitions of $k$ in which no parts occur more than $
s-1 $ times equals  the generating function for the number of
partitions of $k$ such that no parts is divisible by $s$. This is
understood as we have eliminated terms of the form
$\prod_{k=1}^{\infty}(1-x^{sk})$ from the bosonic partition
function which in turn is the generating function for the number
of partitions of $k$ without restrictions. Therefore,  this well
known result in partition theory with restriction is obtained by separating 
the parafermionic Hamiltonians $H_s$, from the bosonic Hamiltonian $H_B$.

 Next, we will give two  explicit examples, the first example is connected with parafermions having non-integer order. The other example that will be considered is connected with the parabosonic of of order $s$, the partner of our parafermion of this section.   
We have seen in the last section that the operators $b^2$ and $(b^{\dag})^2$ are not bosonic operators since the commutator between these operators is not a $c$ number. In this particular situation, it is possible to write the  bosonic Hamiltonian $H=\omega (b^{\dag}b+\frac{1}{2})$ in terms of the scaled operators  $q=\frac{1}{2}(b)^2$ and $q^{\dag}=\frac{1}{2}(b^{\dag})^2$, where $[ q , q^{\dag}]=b{\dag}b+\frac{1}{2}$. Hence, $ H=\omega[ q , q^{\dag}] $. It was shown in \cite{Kademova} that the operators $  q$, $ q^{\dag}$ satisfy a modified version of the Green para-fermi commutation relations  \cite{Green}, $$\frac{1}{2}[ [  q^{\dag},q], q]=-q ,
$$ $$ [ [  q,q], q]=0.$$ The Fock space was constructed whose vacuum $|0\rangle $  is annihilated by $q$, and  the number operator is  $\textbf{n} = \big(\frac{1}{2} [ q, q^{\dag}]-p\big),$ where $p$ is the order of the parafermion defined by $ q q^{\dag}|\emptyset\rangle= p|\emptyset\rangle $, simple computations shows that that the order of parastatistics   are $p= 1/2$, $p= 3/2$. It turns out that the norm of the states in the Fock space of such  infinite number of oscillators is not always positive \cite{Yamada}. For a single oscillator, no such difficulty appear and in this particular case, the bosonic Fock space  is a direct sum of the two Fock spaces associated with parastaistics  $p=1/2$, $p=3/2$ respectively. These Fock spaces are generated by the  states  $$|\textbf{n},p\rangle=(q^{\dag})^{n}|\textbf{n},\emptyset\rangle ,$$ that  are eigenstates of the operator
$$ {\tilde H} = \omega[ q , q^{\dag}]= \omega\big(2\textbf{n}+p\big).$$ 
Now, the eigenstates of $H$ are $|\textbf{n}\rangle = (b^{\dag})^n|\emptyset\rangle$, with the single vacuum $|\emptyset\rangle$, while the states of $ {\tilde H}$ using the realization  $q=\frac{1}{2}(b)^2$,  $q^{\dag}=\frac{1}{2}(b^{\dag})^2$ have two vacuum, $|\emptyset\rangle$ and $b^{\dag}|\emptyset\rangle$, so that $ q q^{\dag}|\emptyset\rangle= 1/2|\emptyset\rangle $, $ q q^{\dag}b^{\dag}|\emptyset\rangle= 3/2|\emptyset\rangle $ . As a consequence, the states using the representation of $H$ for the two Fock spaces may be written as $|\textbf{2n}\rangle $, $n=0,1,\cdots,$ and $|\textbf{2n+1}\rangle $, $n=0,1,\cdots$, respectively. Next, we will write down the expressions for the partition functions in these  sectors,  and show that the sum of these partition functions and  not the  product is identical to the  bosonic partition function. This follows simply by noting that the spectrum in the $p=1/2$ is $E_{n,1/2}= \big(2n+1/2\big)\omega$, while for $p=3/2$ is $E_{n,1/2}= \big(2n+1+1/2\big)\omega$. Therefore their partition functions are; $$ Z_{1/2}(\beta)=\sum_{n=0}^{\infty}\exp(-\beta\omega(2n+{1/2}))=\frac{\exp-(\beta\omega/2)}{1-\exp-(2\beta\omega)}, $$
 and $$ Z_{3/2}(\beta)=\sum_{n=0}^{\infty}\exp(-\beta\omega(2n+1+1/2))=\frac{\exp-(3\beta\omega/2)}{1-\exp-(2\beta\omega)}. $$
 Therefore, the bosonic partition function reads, $$ Z_{B}(\beta)=Z_{1/2}(\beta)+Z_{3/2}(\beta)=\frac{\exp-(\beta\omega)(\exp(\beta\omega/2)+\exp-(\beta\omega/2))}{1-\exp-(\beta\omega)}=\frac{\exp-(\beta\omega/2)}{1-\exp-(2\beta\omega)}.$$
 This shows clearly that $$ Z_{B}(\beta)=Z_{1/2}(\beta)+Z_{3/2}(\beta)=Z_{B}(2\beta)Z_{F}(\beta),$$
 that is,  a bosonic system at temperature $\beta$ is equivalent to a  mixture of a  bosonic system whose temperature is twice as much  and a fermionic system having the same temperature $\beta$. Adding the two parafermionic partition functions at the same temperature may be considered to take place in  the intermediate process.
   \\ In this section, we have seen that the expression for the parafermionic partition function is $  Z_{s}(\beta)=\prod_{k=1}^{\infty}\frac{(1-x^{sk})}{({1-x^k})}$, where $x=\exp(-\beta \omega)$, so that the energy of a given quantum state is a multiple of $\omega$. Suppose we have  a quantum $s$-parafermionic gas system, that is, the maximum number of particles in a given quantum state is $s-1$,  with the total energy  $ E=\sum_{k}n_{k}\epsilon_{k}$, where $k$ labels the one-particle energy levels $ \epsilon_{k}$, and $n_{k}$ is the number of particles  in  state $k$. If $N=\sum_{k}n_{k}$, $N=0,1,2,\cdots $, is the total number of particles and $\mu$ is the chemical potential, Then, the grand canonical partition function may be written as 
 \begin{eqnarray} 
 \label{paraf1}
  Z_{F}^{s}=\prod_{k=1}^{\infty}\frac{(1-x_{k}^{s})}{({1-x_{k}})},
 \end{eqnarray} 
 where $x_{k}=\exp-\beta(\mu-\epsilon_{k})$. One may as well derive  the expression  of the  parabosonic partition function $Z_{B}^{s}$. In the parabosonic case \cite{Safonov}, the occupation number $n$ may take the values $0$, $s, s+1,\cdots$ and the interval of numbers $ 1< n<s$ is not allowed. For $s=1$, the parabosonic gas becomes a bosonic gas. Using the occupation number conditions of the parabosonic gas, then the partition function of a single parabosonic mode follows , $$1+\frac{1}{1-x}-\frac{1-x^s}{1-x}=1+\frac{x^s}{1-x}, $$
 therefore, the expression for parabosonic partition function is
\begin{eqnarray} 
\label{parag1}
    Z_{B}^{s}= \prod_{k=1}^{\infty}\Big(1+\frac{x_{k}^s}{1-x_{k}}\Big).
\end{eqnarray}
If one sets $s=1$ in the last formula, then the bosonic partition function is obtained which is in agreement with the parabose statistics for $s=1$. These results were quoted in \cite{ Hama},  however, the last equation (\ref{parag1}) differs from their expression. Let us consider a parabosonic gas in which the occupation number $n\neq 0$,  and may take the values $s, s+1,\cdots$, then it is clear that the bosonic partition function is the sum  of the parafermionic and parabosonic partition functions. From equation (\ref{parag1}), the partner of the parafermionic partition function is $  \prod_{k=1}^{\infty}\Big(1+\frac{x^{ks}}{1-x^{k}}\Big).$ The quantum states for the above statistics is characterized by specifying  a number of particles in each 1-particle state $|n_{1},n_{2},\cdots\rangle=(a_1^{\dag})^{n_1}(a_2^{\dag})^{n_2}\cdots
(a_k^{\dag})^{n_k}\cdots|\emptyset\rangle$, $ n_{k}\leq s$. Here, all the operators commute among themselves. In the general case the operators do not commute and the Fock spaces of the parabosonic and  the parafermionic systems are described by Green's trilinear relations \cite{Green}. 
Before proving some theorems in the theory of partitions using the
second factorization identity, let us first consider the following
simple example, take a parafermion of order three,  so its partition
function by the second factorization identity is
$Z_{3}(\beta)=\prod_{k=1}^{\infty}(1+ x^{k} +
x^{2k})=\prod_{k=1}^{\infty}\frac{(1-x^{3k})}{({1-x^k})}$, the
right hand side of this identity can be simplified to give
\begin{equation}
\label{toto17}
Z_{3}(\beta)=\prod_{k=1}^{\infty}\frac{1}{(1-x^{3k-2})}\frac{1}{(1-x^{3k-1})},
\end{equation}
where we have used the identity
$\prod_{k=1}^{\infty}(1-x^{3k-2})(1-x^{3k-1})(1-x^{3k})=\prod_{k=1}^{\infty}({1-x^k})$,
the right hand of equation (\ref{toto17}) is the number of
partitions of $k$ into parts prime to 3 and so equals to the
number of partitions of $k$ in which each part occurs at most two
times. Using Maple for the product in equation (\ref{toto17}) one
obtain the following
series\begin{center}
$1+x+2x^{2}+\cdots+9x^{7}+13x^{8}+\cdots+1225x^{30}+\cdots$
.\end{center}
 In terms of partitions this means for example that
the number of partition of the number 7 which are prime to 3 is 9
because
\begin{center}$7=5+2=5+1+1=4+2+1=4+1+1+1=2+2+2+1=2+2+1+1+1=2+1+1+1+1+1=1+1+1+1+1+1+1$.\end{center}
The sequence of partitions numbers corresponds to the coefficients
in the series, therefore, this sequence is
\begin{center}$1,1,2,\cdots,9,13,\cdots,1225,\cdots$,\end{center}
this sequence coincides with the sequence with reference number
$A00726$, in the On-Line encyclopedia of integer sequences
\cite{sloane}. Next, using the second factorization equation (\ref{toto15}), 
we will give a different proof for the following three theorems. one
theorem is on the Andrews' result \cite{andrews} in connection with generating functions that exclude squares and the other two theorems  are their generalizations \cite{sellers}. The Andrews result may be stated  by the following theorem;
\begin{theorem}
 {Let $P_1(n)$ be the number of partitions of $n$ in
which each $k$ appears at most $k-1$ times and let $P_2(n)$ be the
number of partitions of $n$ with no squares part. Then,
$P_1(n)=P_2(n)$}.
\end{theorem}
Proof. The generating function for $P_1(n)$ is nothing but the
parafermionic partition function of order $k$, i.e.,
\begin{equation}
\label{toto18}
\sum_{n=0}^\infty
P_1(n)x^{n}=\prod_{k=1}^{\infty}(1+ x^{k} + x^{2k}+ \cdots +
x^{(k-1)k})=\prod_{k=1}^{\infty}\frac{(1-x^{k^{2}})}{({1-x^k})},
\end{equation}
the right hand side is the generating function for the number of
partitions in which no square parts are present, and so equals to
$\sum_{n=0}^\infty P_2(n)x^{n}$,  and hence $P_1(n)=P_2(n)$. The
generalization considered in \cite{sellers} is to exclude
polygonal numbers or $r$-gons as parts  where the general
polygonal number or the $k^{th} n$-gonal number is given by
$p_k^n=\frac{1}{2}k \left[(n-2)k-(n-4)\right]$, setting $n=4$, this
equation gives square numbers $k^{2}$ and $n=5$ gives the
$k^{th}-pentagonal$ numbers $\frac{1}{2}k (3k-1)$, etc.
\begin{theorem}{Let $r\geq2$ be a fixed integer . Let $P_3(n,r)$ be the number of partitions of $n$
in which each $k$ appears at most $(r-1)(k-1)$ times and let
$P_4(n,r)$ be the number of partitions of $n$ where no $2r$-gons
can be used as parts.Then, $P_3(n,r)=P_4(n,r)$}.
\end{theorem}
Proof. The generating function for $P_3(n,r)$ is the parafermionic
partition function of order $(r-1)(k-1)+1$, and so we have;
\begin{equation}
\label{toto19}
\sum_{n=0}^\infty
P_3(n,r)x^{n}=\prod_{k=1}^{\infty}(1+ x^{k} + x^{2k}+ \cdots +
x^{k(r-1)(k-1)})=\prod_{k=1}^{\infty}\frac{(1-x^{k[(r-1)(k-1)+1]})}{({1-x^k})},
\end{equation}
to complete the proof we simply note that the term
$k[(r-1)(k-1)+1]$ can be rewritten as $k
\left[(r-1)k-(r-2)\right]=\frac{1}{2}k\left[(2r-2)k-(2r-4)\right]$
which is the $2r$-gons, so the right hand side of the above
equation generates partitions whose parts are free of $2r$-gons, 
and hence $P_3(n,r)=P_4(n,r)$. Note that if we set $r=2$, we obtain
the previous result in which no square parts are present in the
partitions.
The next results obtained in \cite{sellers} is to
exclude $2r+1$-gons parts from partitions and is stated as
follows;
\begin{theorem}
{Let $P_5(n,r)$ be the number of partitions of $n$ in
which each the part $2k-1$ $(k\geq1)$ appears at most
$(2r-1)(k-1)$ times (and the frequency of even parts is
unbounded). let $P_6(n,r)$ be the number of partitions of $n$ in
which no odd-subscribed $2r+1$-gons can be used as parts. Then,
for all non-negative $n$ $P_5(n,r)=P_6(n,r)$ }.
\end{theorem}
Proof. The generating function $P_5(n,r)$ from its definition is a
product of even bosonic partition
$\prod_{k=1}^{\infty}\frac{1}{({1-x^{2k}})}$, and the odd
parafermionic partition function of order $[(2r-1)(k-1)+1]$.
Therefore, similarly to the above proof we have
\begin{eqnarray}
\label{toto20}
\sum_{n=0}^\infty
P_{5}(n,r)x^{n}&=&\prod_{k=1}^{\infty}\left(\frac{1}{1-x^{2k}}\right)\left(1+
x^{2k-1} + x^{2(2k-1)}+ \cdots +
x^{(2k-1)(2r-1)(k-1)}\right)\nonumber\\
&=&\prod_{k=1}^{\infty}\left(\frac{1}{1-
x^{2k}}\right)\left(\frac{1- x^{(2k-1)[(2r-1)(k-1)+1]}}{1-
x^{(2k-1)}}\right)\nonumber\\
&=&\prod_{k=1}^{\infty}\left(\frac{1-
x^{(2k-1)[(2r-1)(k-1)+1]}}{1- x^k}\right),
\end{eqnarray}
from the following algebraic identity
\begin{eqnarray}
\label{toto21}
(2k-1)[(2r-1)(k-1)+1]&=&(2k-1)[(2r-1)k-(2r-2)]\nonumber\\
 &=&
\frac{(2r+1-2)}{2}(2k-1)^2-\frac{(2r+1-4)}{2}(2k-1),
\end{eqnarray}
so the term $(2k-1)[(2r-1)(k-1)+1]$ is nothing but the $2r+1$-gons
by definition, therefore $P_5(n,r)=P_6(n,r)$.

\
\section{GRADED PARAFERMIONIC PARTITION FUNCTIONS}
We have already seen in section two that the graded fermionic
partition function $\Delta_F(\beta)={\rm
Tr}\left[(-)^F\exp{-\beta(H_F)}\right]=\prod_{k=1}^{\infty}(1-
x^{k})$, that is,  grading is equivalent to changing the  sign of $x^k$
in the fermionic partition function. Similarly,  the graded bosonic
partition function can be obtained by changing the sign of $x^k$
in the bosonic partition function, this  follow from the fact that the bosonic Fock space of states may be decomposed into direct sum of the Fock space of states of even occupation number and the Fock space of states of odd occupation number . So this space is graded by the operator $(-1)^{N_B}$ so that $(-1)^{N_B}|n_{k}\rangle= (-1)^{n_{k}}|n_{k}\rangle$, that is,  $(-1)^{N_B}$ computes the parity of the  states $ |n_{k}\rangle$ in the  bosonic Fock space of states, it is equal to $\pm1$. Therefore, if $\Delta_B(\beta)$ is the graded bosonic partition function, then 
\begin{eqnarray}
\label{toto22}
\Delta_B(\beta)&=&{\rm
Tr}\left[(-1)^{N_B}\exp{-\beta(H_B)}\right]\nonumber\\&=& \sum_{n_{k}=0}^{\infty}(-1)^{n_{k}} x^{\sum_{k=0}^{\infty}k n_{k}}\nonumber\\&=&\prod_{k=1}^{\infty}\frac{1}{(1+x^k)}.
\end{eqnarray}
Note that the latter operator comes naturally
here in trying to obtain the graded bosonic partition function. It
was introduced in \cite{Spector} in connection with the notion of
the bosonic index; it is $+1$ for Fock space states with an even
number of bosonic creation operators and $-1$ for Fock space
states with an odd number of bosonic creation operators. Recall
from the last section that the parafermion partition function is a
truncation of the bosonic partition function and so the graded
parafermion partition function would be the truncation of the
graded bosonic partition function. There are two truncations to
consider depending on the parity  of the parafermion $(-1)^s=\pm1$. we shall denote
these partition functions by $ \Delta_s^{\pm}(\beta)$. Therefore,  using our physical intuition, the graded parafermionic partition
function is obtained from parafermionic partition function simply
by changing $x^k$ to $-x^k$. Therefore, the identity in equation
(\ref{toto16}) with $x^k$ changed to $-x^k$ gives another
mathematical identity,
\begin{equation}
 \label{toto23}
\prod_{k=1}^{\infty}(1- x^{k} + x^{2k}- \cdots +(-1)^{s-1}
x^{(s-1)k})=\prod_{k=1}^{\infty}\frac{(1+(-1)^{s-1}x^{sk})}{({1+x^k})}.
\end{equation}
Rewriting the above identity in operator form and considering
separately  $s$ even and odd, we end up with the following two
formulae,
\begin{eqnarray}
\label{toto24}
\Delta_s^{+}(\beta)&=&{\rm
 Tr}\left[(-1)^{N_S}\exp{-\beta(H_s)}\right]\nonumber\\
&=&{\rm Tr}\left[(-1)^{N_B}\exp{-\beta(H_B)}\right] {\rm
 Tr}\left[(-1)^F\exp{-\beta(sH_F)}\right],\;\mbox\;s\;\mbox{even}
\end{eqnarray}

\begin{eqnarray}
\label{toto25}
\Delta_s^{-}(\beta)&=&{\rm
Tr}\left[(-1)^{N_S}\exp{-\beta(H_s)}\right]\nonumber\\ &=&{\rm
Tr}\left[(-1)^{N_B}\exp{-\beta(H_B)}\right]{\rm
Tr}\left[\exp{-\beta(sH_F)}\right],\;\mbox\;s\;\mbox{odd}
\end{eqnarray}
where the operator $(-1)^{N_S}$ is like $(-1)^{N_B}$ it is $+1$ for
Fock space states with an even number of parafermions  and $-1$ for Fock space states with an odd number of parafermions. The above identities are identical
to equations (6.1) and (6.2) given in \cite{Spector}. Next,  We will
seek for the physical meaning associated with the generating
functions $ \Delta_s^{\pm}(\beta)$. For $s=2$, equation
(\ref{toto23})  gives
$1/\Delta_2^{+}(\beta)=\prod_{k=1}^{\infty}\frac{({1+x^k})}{(1-x^{2k})}$,
which is the bosonic partition function. This can be written as
$Z_b(\beta)=Z_b(2\beta)Z_f(\beta)=\prod_{k=1}^{\infty}\frac{1}{({1-x^k})}$,
that is, the partition function of the Euler gas. This partition
function is known to be the generating function of partitions
without restrictions;
$\prod_{k=1}^{\infty}\frac{1}{({1-x^k})}=\sum_{k=0}^{\infty}p(k)x^k$,
where $p(k)$ is the number of unrestricted partitions. One may obtain the same results by considering $s=2$ parafermion, the partition function is $\prod_{k=1}^{\infty}{({1+x^k})}$, and so the inverse of the graded parafermion partition function (IGPPF) is $1/\Delta_{F}(\beta)=\prod_{k=1}^{\infty}\frac{1}{(1-x^{k})}$ the bosonic partition function (infinite order parafermionic parition function).  Similarly, for boson the partition function (IGPPF) is the fermionic partition function $\prod_{k=1}^{\infty}{({1+x^k})}$.  Setting $s=1$,
gives
$1/\Delta_1^{-}(\beta)=\prod_{k=1}^{\infty}\frac{({1+x^k})}{(1+x^{k})}=1$,
this means we have a complete cancellation of fermions. Therefore, to
make connection with partition theory or physics, the generating
functions that we should  consider is $1/ \Delta_s^{\pm}(\beta)$ depending on the parity of $s$, and
so using equation (\ref{toto23}), the  partition function (s-IGPPF) may be written as
\begin{equation}
\label{toto26}
\frac{1}{\Delta_s^{+}(\beta)}=Z_B(s\beta)Z_F(\beta),
\end{equation}

\begin{equation}
\label{toto27}
\frac{1}{\Delta_s^{-}(\beta)}=Z_F(\beta)\Delta_B(s\beta),
\end{equation}
as a consequence, $\frac{1}{\Delta_s^{+}(\beta)}$ for $s=4,6,8,\cdots,$ is a
bosonic partition function with some bosonic states missing, due to
the partition function $Z_B(s\beta)$. Similarly,
$\frac{1}{\Delta_s^{-}(\beta)}$ for $s=3,5,7,\cdots,$ is a fermionic
partition function with some fermionic states missing due to the
graded bosonic partition function $\Delta_B(s\beta)$. For $s$ odd one may write the partition function as $$\frac{1}{\Delta_s^{-}(\beta)}= \prod_{k=1,k\nmid{s}}^{\infty}(1+x^k)=\prod_{k=1,k\nmid{s}}^{\infty}\frac{1}{(1-x^{2k-1})}.$$ For example, consider a non-relativistic bosonic quantum string \cite{zwiebach} 
whose Hamiltonian $H_B=\omega\sum_{k=1}^\infty k a_k^{\dag}a_k$ that
is, the $k^{th}$ harmonic oscillator has a frequency $k\omega$. The operators $a_k$ and $a_k^{\dag}$
satisfy  the usual commutators given by $[ a_m , a_n^{\dag} ]
=\delta_{nm}$.  The ground state $|\emptyset\rangle$ is defined by
$a_k |\emptyset\rangle=0$, for all $k$ . The quantum states of the
string denoted by $|n\rangle$ are obtained by acting with the
creation operators on the ground state,i.e,
$|n\rangle=(a_1^{\dag})^{n_1}(a_2^{\dag})^{n_2}\cdots
+(a_k^{\dag})^{n_k}\cdots|\emptyset\rangle$. Applying the number
operator $\textbf{n}=\sum_{k=1}^\infty k a_k^{\dag}a_k$ on the state
$|n\rangle$, we have $\textbf{n}|n\rangle=n |n\rangle$ where
$n=\sum_{k=1}^\infty kn_k$. One can see that for a fixed $n$, the
solutions to the latter equation are given by $p(n)$. Therefore
$p(n)$ counts the number of states with a given eigenvalue $n$. for
example $n=3$, $p(3)=3$ since the partitions of $3$ are; $3$, $2+1$,
and $1+1+1$. In terms of quantum states with a number eigenvalue
$3$, we have $3$ states which are $a_3^{\dag}|\emptyset\rangle$,
$a_2^{\dag}a_1^{\dag}|\emptyset\rangle$ and
$(a_1^{\dag})^3|\emptyset\rangle$. Therefore, a Quantum state is
constructed by attaching each part of the partitions  a subscript
to an oscillator $a^{\dag}$. In the same way one may consider a
fermionic non-relativistic quantum string whose Hamiltonian is
$H_F=\omega\sum_{k=1}^\infty k\psi _k^{\dag}\psi_k$ with $\{ \psi_m
, \psi_n^{\dag} \} =\delta_{nm}$. The same construction follows
for the quantum fermionic states in which we attach each part of the
partitions to an oscillator $\psi^{\dag}$, with the constraint that
the configuration numbers, $n_k=1$ or $0$. Note that the Euler
theorem in terms of quantum states is equivalent to say for a fixed
eigenvalue number $n$, the number of fermionic states  is equal to
the number of bosonic states obtained from bosonic creation
operators with odd subscripts, and with the same eigenvalue number
$n$. Therefore,  fermionic states are parts of the bosonic states in
the Euler gas. Now, let us consider the graded parafermion of order
three, its partition function is
$\Delta_3^{-}(\beta)=\prod_{k=1}^{\infty}(1- x^{k} +
x^{2k})=\prod_{k=1}^{\infty}\frac{(1+x^{3k})}{({1+x^k})}$, or equivalently
\begin{eqnarray}
\label{toto28}
\Delta_{3}^{-}(\beta)&=&\prod_{k=1}^{\infty}(1- x^{k}+
x^{2k})\nonumber\\
&=&\prod_{k=1}^{\infty}\frac{(1+x^{3k})}{({1+x^k})}\nonumber\\
&=&\prod_{k=1}^{\infty}\frac{1}{(1+x^{3k-2})}\frac{1}{(1+x^{3k-1})}.
\end{eqnarray}
Then the partition function (3-IGPPF)  may be written as 
\begin{eqnarray}
\label{toto29}
\frac{1}{\Delta_{3}^{-}(\beta)}&=&\prod_{k=1}^{\infty}{(1+x^{3k-2})}{(1+x^{3k-1})}\nonumber\\&=&\prod_{k=1}^{\infty}\frac{(1-x^{3(2k-1)})}{(1-x^{2k-1})}\nonumber\\&=&
\prod_{k=1}^{\infty}\frac{1}{(1-x^{6k-5})}\frac{1}{(1-x^{6k-1})},
\end{eqnarray}
where the first line of the  above equation is  the generating
function of the number of partitions of $k$ into distinct parts
which are prime to 3. The last line  On the hand represents the generating function for the number of  partitions of  $k$ into
parts congruent to 1 or 5 modulo 6. This was proved by I. Schur In 1926 \cite{schur}.
Thus,  we learn that the partition function (3-IGPPF) coincides with the generating function for the
numbers of partitions in the Schur's theorem. In terms of quantum
states, the only fermionic operators $ \psi_k^{\dag}$ that we
consider are those for which $k$ is congruent to $ 1, 2 $ modulo
$3$. When $s=4$, then, the partition function (4-IGPPF) is
\begin{eqnarray}
\label{toto30}
\frac{1}{\Delta_{4}^{+}(\beta)}&=&
\prod_{k=1}^{\infty}\frac{(1+x^{k})}{(1-x^{2(2k)})}\nonumber\\
&=&\prod_{k=1}^{\infty}\frac{(1+x^{2k-1})}{(1-x^{2k})},
\end{eqnarray}
where the the first product on the right hand side of this
equation gives the number of partitions of $k$ in which each even
part occurs with even multiplicity. There is no restriction on the
odd parts. This generating function corresponds to sequence A006950
\cite{sloane}. It is also the number of partitions of $k$ in which
all odd parts occur with multiplicity one. There is no restriction
on the even parts. This was a comment that we made on this
sequence \cite{sloane}. In terms of quantum states, equation
(\ref{toto30}), tells us that these states are constructed such
that the operators $( a_k^{\dag})$, $k$ is a multiple of 2, appear
with even multiplicity. There is no restriction on those operators
whose subscripts are odd. For example the number of partitions of
$4$ is $3$ because $4$ can be written as $2+2$, $3+1$ and
$1+1+1+1$. The corresponding quantum states are
$a_2^{\dag^2}|\emptyset\rangle$,
$a_3^{\dag}a_1^{\dag}|\emptyset\rangle$ and
$(a_1^{\dag})^4|\emptyset\rangle$. It was shown in  \cite{Vafa}, that the  partition function given in equation (\ref{toto30}), is related to the irreducible  characters  $ \chi_{0}(q )$ and  $ \chi_{1/2}(q )$ of the $c=1/2$ Virasoro algebra. More explicitly, the coefficients in $$q^{-1/16}\frac{\chi_{0}(q )}{\eta}=q^{-1/16}\sum_{n=0}^{\infty}b_{n}q^{n}=q^{-1/16}(1+q+3q^{2}+5q^{3}+10q^{4}+\cdots),$$ and   $$q^{7/16} \frac{\chi_{1/2}(q )}{\eta}=q^{7/16}\sum_{n=0}^{\infty}c_{n}q^{n}=q^{71/16}(1+2q+4q^{2}+7q^{3}+13q^{4}+\cdots),$$  are given by $b_{n}=a_{2n}$, $c_{n}=a_{2n+1}$, respectively, where $a_{n}$ is the expansion coefficient of $ \frac{1}{\Delta_{4}^{+}(\beta)}$. As a consequence, the coefficients $ b_{n}$, $c_{n}$ may be given interpretation in terms of partitions, this follows easily from the above definitions and it is not difficult to see that $b_{n}$ is the number of partitions of $2n$, such that  even part occurs with even multiplicity. There is no restriction on the
odd parts. Alternatively, the number of partitions of $2n$ in which
all odd parts occur with multiplicity one. There is no restriction
on the even parts. The same definition is given to $c_{n}$, but for the partition of the number $2n+1$. Very recently, the coefficients of the partition function (4-IGPPF), are associated  with the counting  of the Moore-Read edge spectra  in the anti-periodic sectors \cite{Michael}. The next generating functions
to consider, are the partition functions (5-IGPPF) and (6-IGPPF) respectively,
\begin{eqnarray}
\label{toto31}
\frac{1}{\Delta_{5}^{+}(\beta)}&=&
\prod_{k=1}^{\infty}\frac{(1+x^{k})}{(1+x^{5k})}\nonumber\\&=&
\prod_{k=1}^{\infty}(1+x^{5k-1})(1+x^{5k-2})(1+x^{5k-3})(1+x^{5k-4})\nonumber\\
&=&\prod_{k=1}^{\infty}\frac{1}{(1-x^{10k-1})(1-x^{10k-3})(1-x^{10k-7})(1-x^{10k-9})},
\end{eqnarray}

\begin{eqnarray}
\label{toto32}
\frac{1}{\Delta_{6}^{+}(\beta)}&=&
\prod_{k=1}^{\infty}\frac{(1+x^{k})}{(1-x^{3(2k)})}\nonumber\\
&=&\prod_{k=1}^{\infty}\frac{(1+x^{3k-1})(1+x^{3k-2})}{(1-x^{3k})}\nonumber\\
&=&\prod_{k=1}^{\infty}\frac{1}{(1-x^{6k-1})(1-x^{6k-3})(1+x^{6k-5})(1-x^{6k})}.
\end{eqnarray}
Equation (\ref{toto31}) gives the number of partitions of $k$ into
distinct parts prime to 5. Equivalently, It is the number of
partitions of $k$ into parts congruent to $ 1,3,7,9 $ modulo $10 $.
This generalizes parts of Schur's theorem. The second definition is
seen simply by applying the Euler theorem to the fourth term on the
right hand side of equation (\ref{toto31}), then write each
denominator as a product of even and odd parts. The first term of
the right hand side of equation (\ref{toto31}), gives the number of
partitions of $k$ in which each even part occurs with a multiple of
3. There is no restriction on the odd parts. From the second term of
equation (\ref{toto31}), it is also the number of partitions into
parts congruent to $ 0, 1, 3, 5$ modulo $6 $. The generating
functions $\frac{1}{\Delta_{5}^{-}(\beta)}$,
$\frac{1}{\Delta_{6}^{+}(\beta)}$ correspond to  sequences by the author
A096938 and A096981 respectively, \cite{sloane}. Let us now consider
some examples associated with the partition functions $1/
\Delta_s^{\pm}(\beta)$. Using Maple,  one could generate sequences of
partitions from our formulae depending on $s$. When $s=3$ we have,
\begin{center}$1+x+x^{2}+\cdots+3x^{7}+3x^{8}+3x^{9}+4x^{10}+\cdots $\end{center}
so for example the number of distinct partitions of the number 10 in
which the parts are prime to 3 is 4 (which we denote by $a(10)=4$)
because $10=8+2=7+2+1=5+4+1$. The corresponding sequence of
partitions is,\begin{center} $
1,1,1,1,1,2,2,3,3,3,4,5,6,7,8,9,10,12,\cdots $,\end{center} this is
the sequence $A003105$ on the on-Line Encyclopedia of integer
sequences \cite{sloane}. From Schur theorem the number of partitions
10 prime to 3 is equal to the number of partitions of 10 congruent
to $1, 5$ modulo $10$. This is so, as 10 can be written as $
7+1+1+1=5+5=5+1+1+1+1+1=1+1+1+1+1+1+1+1+1+1 $. The second part of
Schur's theorem tells us that both of these number of partitions are
equal to the partition of $10$ into parts  that differs by at least
$3$. This is indeed the case, because we can write $10$ as, $
10=9+1=8+2=7+3$. In terms of quantum states of the quantum string,
the number of fermionic states in which the subscripts of the
fermionic operators are congruent to $ 1, 2 $ modulo $3$, equals the
number of bosonic states in which the subscripts of the bosonic
operators are congruent to $ 1,5 $ modulo $6$. The next sequence of
partitions of the number $k$ would be that which corresponds to
$s=4$, which is,\begin{center} $1,1,1,2,3,4,5,7,10,13,16,21,\cdots
$\end{center} this is also in \cite{sloane} and corresponds to the
sequence A006950. It was pointed out by Sellers, that the number of
partitions  for this sequence that we mentioned above for $s=4$, it
is also the number of partitions of $k$ into parts not congruent to
2 mod 4. therefore for $s=4$, we have three definitions. For
example, the number of partition of 7 equals 7, $a(7)=7$ because
\begin{center}$7=5+1+1=3+3+1=3+1+1+1+1=3+2+2=2+2+1+1+1=1+1+1+1+1+1+1$,\end{center}
\begin{center}$7=6+1=5+2=4+3=4+2+1=3+2+2=2+2+2+1$.\end{center} or
\begin{center}$7=5+1+1=4+3=4+1+1+1=3+3+1=3+1+1+1+1=1+1+1+1+1+1+1$.\end{center}
The fifth and the sixth orders from our formulae for the inverted
parafermionic partition function give the following sequences of
partition of the number $k$,
\begin{center}
$1,1,1,2,2,2,3,4,4,6,7,8,10,16.19,22,26,30,35,41,\cdots$\end{center}
for the fifth order and
\begin{center}$1,1,1,2,2,3,5,6,7,10,12,15,25,30,39,46,\cdots$\end{center}
for the sixth order respectively. In the fifth order, of  the inverted
parafermionic partition function, the number 5 and its multiples are
not present in the  partitions of $k$ into distinct parts, for
example $9=8+1=7+2=6+3=6+2+1=4+3+2$ so $a(9)=6$. From the second
definition for $s=5$, this also the number of partition of 9 into
parts congruent to $1, 3, 7, 9$ modulo $10$. This is indeed the
case, since
$9=7+1+1=3+3+3=3+3+1+1+1=3+1+1+1+1+1+1=1+1+1+1+1+1+1+1+1$. For the
sixth order sequence the number of partitions of the number 11 for
example, is 15, $a(11)=15$. Because
\begin{center}$
11=9+1+1=7+3+1=7+1+1+1+1=5+5+1=5+3+3=5+3+1+1+1=5+2+2+2=5+1+1+1+1+1+1=3+3+3+1+1
=3+3+1+1+1+1+1=3+2+2+2+1+1=3+1+1+1+1+1+1+1+1=2+2+2+1+1+1+1+1=1+1+1+1+1+1+1+1+1+1+1.$\end{center}
This is also given by the number of partitions of 11 congruent to $
0, 1, 3, 5$ modulo $6 $. Because
\begin{center}$
11=9+1+1=7+3+1=7+1+1+1+1=6+5=6+3+1+1=6+1+1+1+1+1=5+5+1=5+3+3=5+3+1+1+1=5+1+1+1+1+1+1=3+3+3+1+1
=3+3+1+1+1+1+1=3+1+1+1+1+1+1+1+1=1+1+1+1+1+1+1+1+1+1+1.$\end{center}
Therefore, we see that our generating function gives known sequences
for partitions of integers as well as new ones. In terms of
partition functions, when $s$ is even then, the generating function
$\frac{1}{\Delta_s^{+}(\beta)}$ is definitely bosonic partition with
some restrictions. This partition function is identical to an other
bosonic partition function with different restrictions. When $s$ is
odd, then, the generating function $\frac{1}{\Delta_s^{-}(\beta)}$ is
a fermionic partition function, for  for $s=1$, we have a complete cancelation of fermions . One should note, however, that
the fermionic partition function is identical to a bosonic partition
function with different restrictions, i.e, we have a fermi-bose
equivalence. From the explicit examples given above for $s=3, 4, 5$
and $6$, one can see that the definitions for
$\frac{1}{\Delta_s^{\pm}(\beta)}$ can be generalized for any $s$,
and are summarized by the following proposition.
\begin{proposition} The bosonic partition function
$\frac{1}{\Delta_s^{+}(\beta)}$, is the generating function for
partitions of $k$ in which all odd parts are unrestricted. The even
parts occur with multiples of $s/2$. Alternatively it is the number
of partitions of $k$ into parts congruent to $ 0, 1, 3,
5,\cdots,(s-1)$ modulo $s$. The fermionic partition function,
$\frac{1}{\Delta_s^{-}(\beta)}$, is the generating function for
partitions of $k$ into distinct parts prime to $s$. This is also the
generating function for partitions of $k$ into parts congruent to $
1, 3, 5,\cdots,(\hat{s})\cdots,(2s-1)$ modulo $(2s)$. Note that
$\hat{s}$ means that the parts are not congruent to $s$ modulo
$(2s)$.
\end{proposition}
 Proof:
\begin{eqnarray}
\label{toto33}
\frac{1}{\Delta_s^{+}(\beta)}&=&
\prod_{k=1}^{\infty}\frac{(1+x^{k})}{(1-x^{sk})} \nonumber\\
&=&\prod_{k=1}^{\infty}\frac{1}{(1-x^{2k-1})(1-x^{s/2(2k)})} \nonumber\\
&=&\prod_{k=1}^{\infty}\frac{1}{(1-x^{sk})(1-x^{sk-1})(1-x^{sk-3})\cdots(1-x^{sk-(s-1)})},
\end{eqnarray}
where the last line, may be obtained  from the following identity
\begin{center}$\prod_{k=1}^{\infty}{(1-x^{2k-1})=
\prod_{k=1}^{\infty}(1-x^{sk-1})(1-x^{sk-3})\cdots(1-x^{sk-(s-1)})}$,\end{center}
for $s$ even. This proves the first part of the proposition. Using the
same identity for $s$ odd, and change $s$ to $2s$ and the Euler
identity, then
\begin{eqnarray}
\label{toto34}
\frac{1}{\Delta_s^{-}(\beta)}&=&
\prod_{k=1}^{\infty}\frac{(1+x^{k})}{(1+x^{sk})} \nonumber\\
&=&\prod_{k=1}^{\infty}\frac{1}{(1-x^{2k-1})(1+x^{sk})} \nonumber\\
&=&\prod_{k=1}^{\infty}\frac{(1-x^{2sk-s})}{(1-x^{2sk-1})(1-x^{2sk-3})
\cdots(1-x^{2sk-s})\cdots(1-x^{2sk-(2s-1)})}
\end{eqnarray}
the last identity shows that the fermionic partition is indeed the
generating function for partitions of $k$ into parts congruent to $
1, 3, 5,\cdots,(\hat{s})\cdots,(2s-1)$ modulo $(2s)$. Being a
generating function of partitions into distinct parts prime to $s$ is
obvious from the first identity. Before finishing
this section,  we would like to point out that the generating
functions $\frac{1}{\Delta_s^{\pm}(\beta)}$ may be considered as
truncation of fermions. They are the analogue of of truncation
of bosons. Explicitly,
\begin{eqnarray}
\label{toto35}
\frac{1}{\Delta_s^{+}(\beta)}&=&\prod_{k=1}^{\infty}\frac{(1+x^{k})}
{(1-x^{(s/2)k)})(1+x^{(s/2)k})}\nonumber\\
&=&\prod_{k=1,k\nmid{s/2}}^{\infty}(1+x^{k})\prod_{k=1}^{\infty}\frac{1}
{(1-x^{(s/2)k})},
\end{eqnarray}
where the notation $k\nmid{s/2}$, means that $k$ is not divisible by
$s/2$. Here,we have a sort of exchange between fermions and bosons, that is, those fermion missing from the spectrum are replaced by bosons. Therefore for $s$ even, the only fermionic operators that
appear in the construction of the fermionic states, are
$\psi_k^{\dag}$ for which $k$ is not divisible by $s/2$. For the
bosonic operators, however, $a_k^{\dag}$ are those for which $k$ is
a multiple of $s/2$. when $s$ is odd we have a fermionic truncation,
\begin{eqnarray}
\label{toto36}
\frac{1}{\Delta_s^{-}(\beta)}&=&\prod_{k=1}^{\infty}\frac{(1+x^{k})}
{(1+x^{sk})}\nonumber\\
&=&\prod_{k=1,k\nmid{s}}^{\infty}(1+x^{k}).
\end{eqnarray}

\section{GENERATING FUNCTIONS AND THE JACOBI THETA-FUNCTION, $\theta_4(0,x)$}
In this section, we will construct partition functions from the
parafermionic partition function and the inverse graded parafermion partition
function (s-IGPPF). We will find that our partition functions are connected
with the Jacobi theta function $\theta_4(0,x)$, and also find some
sequences in partition theory. These partition functions
corresponding to two different systems, namely, either by mixing
parafermionic system of order $s$ with a system whose partition
function is $\frac{1}{\Delta_s^{+}(\beta)}$ or
$\frac{1}{\Delta_s^{-}(\beta)}$. Therefore,  using the identities
given in equations (\ref{toto16}), (\ref{toto26}), and
(\ref{toto27}) we obtain the following two partition functions;
\begin{eqnarray}
\label{toto37}
\frac{Z_{s}(\beta)}{\Delta_s^{+}(\beta)}&=&
Z_{F}(\beta)Z_{B}(\beta)\nonumber\\
&=&\frac{\prod_{k=1}^{\infty}(1+
x^{k} + x^{2k}+ \cdots + x^{(s-1)k})}{\prod_{k=1}^{\infty}(1- x^{k}
+ x^{2k}- \cdots -
x^{(s-1)k})}\nonumber\\
&=&\prod_{k=1}^{\infty}\frac{(1+x^k)}{(1-x^k)},
\end{eqnarray}

\begin{eqnarray}
\label{toto38}
\frac{Z_{s}(\beta)}{\Delta_s^{-}(\beta)}&=&
Z_{F}(\beta)Z_{B}(\beta)\Delta_{B}(s\beta)\Delta_{F}(s\beta)\nonumber\\
&=&\frac{\prod_{k=1}^{\infty}(1+ x^{k} + x^{2k}+ \cdots +
x^{(s-1)k})}{\prod_{k=1}^{\infty}(1- x^{k} + x^{2k}- \cdots +
x^{(s-1)k})}\nonumber\\
&=&\prod_{k=1}^{\infty}\frac{(1+x^k)}{({1-x^k})}\frac{(1-x^{sk})}{(1+x^{sk})}.
\end{eqnarray}
Here, we would like to make some remarks  about the above identities
and their implications. The first identity shows that for  $s$
even, mixing a parafermionic system of order $s$ with a system whose
partition function is $\frac{1}{\Delta_s^{+}(\beta)}$, is identical to a mixed system of  bosons  and fermions. Note the independence on the parameter  $ s $ of the partition function of the mixed system in this case.  This follows simply  from the the cancellation of the expression $Z_{B}(s\beta )$ from the partition functions  $ {Z_{s}(\beta)}$ and  $\frac{1}{\Delta_s^{+}(\beta)}$. Therefore, for $s$ even,  the same expression for the total partition function of the mixed system may be obtained by letting the order $s$ in both the parafermion partition function and the partition function (s-IGPPF) to go to infinity.  For $s$ odd, there are
partial cancellations between $Z_{F}$ and $\Delta_{B}(s\beta)$,
similarly between $Z_{B}(\beta)$ and $\Delta_{F}(s\beta)$. Due to the cancellations of bosons and fermions in this case,  we may as well write equation (\ref{toto38}) in the  form $$ \frac{Z_{s}(\beta)}{\Delta_s^{-}(\beta)}=\prod_{k=1,k\nmid{s}}^{\infty}\frac{(1+x^k)}{({1-x^k})}.$$
Mathematically speaking, for $s$ even the ratio of the parafermionic
partition function to the graded parafermionic partition function is
always given by $\prod_{k=1}^{\infty}\frac{(1+x^k)}{({1-x^k})}$. This product may be written as a sum if we set $a=1$ in the following formula \cite{corteel} $$ \sum_{n=0}^\infty\frac{(-a;x)_{n}}{(x;x)_{n}}x^{n}=\frac{(-ax;x)_{\infty}}{(x;x)_{\infty}},$$  where  $(a;x)_{n}:=(1-a)(1-ax)\cdots(1-ax^{n-1}) $, $(a;x)_{0}=1$ and $(a;x)_{\infty}:=\prod_{n=0}^{\infty}(1-ax^{n})$, to obtain
\begin{eqnarray}
\label{toto39}
\frac{Z_{s}(\beta)}{\Delta_s^{+}(\beta)}
&=&\prod_{k=1}^{\infty}\frac{(1+x^k)}{(1-x^k)} \nonumber\\&=&
1+\sum_{k=1}^\infty\frac{2(1+x)(1+x^{2})\cdots(1+x^{k-1})}
{(1-x)(1-x^{2})\cdots(1-x^{k})}x^k.
\end{eqnarray}
This is known to be equal to the inverse of
$\theta_4(0,x)=\sum_{n=-\infty}^\infty (-1)^n x^{n^{2}}$ via
Gauss's identity. The second identity, in the case of $s$ odd,
does not give $\frac{1}{\theta_4(0,x)}$ but rather the ratio
$\theta_4(0,x^s)/\theta_4(0,x)$, so this extra factor can be
thought of as correction factor that one has to insert in
$\frac{Z_{s}(\beta)}{ \Delta_s^{+}(\beta)}$. Now, rewriting our
partition functions in terms of $\theta_4(0,x)$, then we obtain;
\begin{eqnarray}
\label{toto40}
\frac{Z_{s}(\beta)}{\Delta_s^{+}(\beta)} &=&
\prod_{k=1}^{\infty}\frac{(1+x^{k})}{(1-x^{k})}\nonumber\\
&=&\frac{1}{\theta_4(0,x)},\;\mbox\;s\;\mbox{even}
\end{eqnarray}

\begin{eqnarray}
\label{toto41}
\frac{Z_{s}(\beta)}{\Delta_s^{-}(\beta)} &=&
\prod_{k=1}^{\infty}\frac{(1+x^{k})}{(1-x^{k})}\frac{(1-x^{sk})}{(1+x^{sk})}\nonumber\\
&=&\frac{\theta_4(0,x^s)}{\theta_4(0,x)},\;\mbox\;s\;\mbox{odd}
\end{eqnarray}
 mixing a parafermionic system of order $s$ with a
system whose partition function, $\frac{1}{\Delta_s^{\pm}(\beta)}$,
we produce two systems whose partition functions are 
$\frac{1}{\theta_4(0,x)}$ and
$\frac{\theta_4(0,x^s)}{\theta_4(0,x)}$ respectively. The second partition
function is obtained from the first one by canceling some of the
states from the full Fock space of bosons and fermions. Note that the  partition $\frac{1}{\theta_4(0,x)}$, is also the partition function for the free theory formed by a single holomorphic boson and holomorphic fermion with anti-periodic spin structure on the torus \cite{Vafa}. In terms of
partition theory, and referring to sequence A015128 in
\cite{sloane}, $\frac{1}{\theta_4(0,x)}$ is known to be generating
function for partitions of $2k$ with all odd parts occurring with
even multiplicities. There is no restriction on the even parts. The
proof is straightforward, and follows from the Euler theorem;
\begin{eqnarray}
\label{toto42}
\frac{1}{\theta_4(0,x)}&=&
\prod_{k=1}^{\infty}\frac{(1+x^{k})}{(1-x^{k})}, \nonumber\\
&=&\prod_{k=1}^{\infty}\frac{1}{(1-x^{(2k-1)})(1-x^{k})},
\end{eqnarray}
then by changing the variable $x$ to $x^2$, and using equation
(\ref{toto35}), it follows that the expansion coefficients of
$\frac{1}{\theta_4(0,x^2)}$ are identical to those of
$\frac{1}{\theta_4(0,x)}$. Therefore, the generating function
$\frac{1}{\theta_4(0,x^2)}=\prod_{k=1}^{\infty}\frac{1}{(1-x^{2(2k-1)})(1-x^{2k})}$
says that; the odd parts occur with even multiplicities. There is
no restriction on the even parts. This is  the definition we have given to
the sequence A015128. Using Maple, the sequence for
$\frac{1}{\theta_4(0,x)}$ is,
\begin{center}$1,2,4,8,14,24,40,64,100,154,\cdots,9904,13288,\cdots.$\end{center}
This sequence also appear  in the counting of the Moore-read edge spectra together with the sequence generated by the partition function (4-IGPPF), however, the above sequence is associated  with a periodic sector \cite{Michael}.
For example, the number of partitions of $6$ is $8$. This is
because, we can write $6$ as
\begin{center}$6=4+2=4+1+1=3+3=2+2+2=2+2+1+1=2+1+1+1+1=1+1+1+1+1+1$.\end{center}
In terms of quantum states, for a fixed number eigenvalue $n$, the
number of states obtained from the total partition function given by
equation (\ref{toto40}) should be equal to the number of bosonic
states with a fixed number eigenvalue $2n$. The states are
constructed by applying creation operators $a_k^{\dag}$ on the
ground state whose subscripts are the parts of the partitions of
$2n$. We can check this for the above example as follows, the Fock
space of states corresponding to the partition function given by
equation (\ref{toto40}) is a tensor product of the fermionic and
bosonic Fock spaces. The bosonic Fock space of states having a
number eigenvalue up to $3$ are;
\begin{center}$ a_{3}^{\dag}|\emptyset\rangle,
a_{2}^{\dag}a_1^{\dag}|\emptyset\rangle,
a_{2}^{\dag}|\emptyset\rangle, a_{1}^{\dag^2}|\emptyset\rangle,
a_{1}^{\dag}|\emptyset\rangle, a_{1}^{\dag^3}|\emptyset\rangle,
|\emptyset\rangle $.\end{center} For the fermionic Fock space of
states whose ground state is $|\tilde{\emptyset}\rangle$, we have
\begin{center} $\psi{_3}^{\dag}|\tilde{\emptyset}\rangle,
\psi_{2}^{\dag}\psi_{1}^{\dag}|\tilde{\emptyset}\rangle,
|\tilde{\emptyset}\rangle, \psi_{2}^{\dag}|\tilde{\emptyset}\rangle,
\psi_{1}^{\dag}|\tilde{\emptyset}\rangle,|\tilde{\emptyset}\rangle
$.\end{center} Now taking the tensor product of these states such
that $n=3$, we obtain the following states;
\begin{center}$
\psi_{3}^{\dag}|\tilde{\emptyset}\rangle\otimes|\emptyset\rangle,
\psi_{2}^{\dag}\psi_{1}^{\dag}|\tilde{\emptyset}\rangle\otimes|\emptyset\rangle,
\psi_{2}^{\dag}|\tilde{\emptyset}\rangle\otimes
a_{1}^{\dag}|\emptyset\rangle,
\psi_{1}^{\dag}|\tilde{\emptyset}\rangle\otimes
a_{2}^{\dag}|\emptyset\rangle,
\psi_{1}^{\dag}|\tilde{\emptyset}\rangle\otimes
a_{1}^{\dag^2}|\emptyset\rangle, |\tilde{\emptyset}\rangle\otimes
a_{3}^{\dag}|\emptyset\rangle, |\tilde{\emptyset}\rangle\otimes
a_{2}^{\dag}a_1^{\dag}|\emptyset\rangle,
|\tilde{\emptyset}\rangle\otimes
a_1^{\dag^3}|\emptyset\rangle$.\end{center}  These are exactly the
number of partitions of $6$. We have seen that mixing a
parafermionic system of an even order, with a system whose partition
function  (s-IGPPF) having the same
order gives a bosonic system and has a number
eigenvalue $2n$. These states for the number eigenvalue $6$, are;
\begin{center}$ a_{6}^{\dag}|\emptyset\rangle,
a_{4}^{\dag}a_{2}^{\dag}|\emptyset\rangle,
a_{4}^{\dag}a_{1}^{\dag^2}|\emptyset\rangle
a_{3}^{\dag^2}|\emptyset\rangle, a_{2}^{\dag^3}|\emptyset\rangle,
a_{2}^{\dag^2}a_{1}^{\dag^2}|\emptyset\rangle,
a_{2}^{\dag}a_{1}^{\dag^4}|\emptyset\rangle,
a_{1}^{\dag^6}|\emptyset\rangle $.\end{center} In the case of a
system whose partition function is
$\frac{\theta_4(0,x^s)}{\theta_4(0,x)}$ for $s$ odd, it turns out
that this particular partition function generates  sequences in
partition theory. As an example setting $s=3$ in equation
(\ref{toto41}), then we have
\begin{equation}
\label{toto43}
\frac{Z_{3}(\beta)}{\Delta_3^{-}(\beta)}=
\frac{\theta_4(0,x^3)}{\theta_4(0,x)},
\end{equation}
from the product-sum identity $6$ in Slater list on the
Rogers-Ramanujan type identities \cite{slater}, corrected by Sills
\cite{sills} in a Ph.D thesis, we learn that our partition function
can be written by the following explicit sum;
\begin{eqnarray}
\label{toto44}
\frac{\theta_4(0,x^3)}{\theta_4(0,x)}&=&
\prod_{k=1}^{\infty}\frac{(1+x^{3k-1})(1+x^{3k-2})}{(1-x^{3k-1})(1-x^{3k-2})}\nonumber\\
&=& 1+\sum_{k=1}^\infty\frac{2(1+x)(1+x^{2})\cdots(1+x^{k-1})}
{(1-x)(1-x^{2})\cdots(1-x^{k})(1-x)(1-x^{3})\cdots(1-x^{2k-1})}x^{k^{2}}.
\end{eqnarray}

Using Maple the sequence corresponding to
$\frac{\theta_4(0,x^3)}{\theta_4(0,x)}$ is,
\begin{center}$1,2,4,6,10,16,24,36,52,74,104,144,198,268,360,\cdots,48672,59122,\cdots$.\end{center}
This corresponds to the number of partitions of $2k$ prime to 3
with all odd parts occurring with even multiplicities. There is no
restriction on the even parts e.g $a(6)=6$ because we can write
$6$ as,
\begin{center}
$4+2=4+1+1=2+2+2=2+2+1+1=2+1+1+1+1=1+1+1+1+1+1.$\end{center} This sequence has reference number A098151 on the on-Line
Encyclopedia of integer sequences \cite{sloane}. This can be proved
as follows;
\begin{eqnarray}
\label{toto45}
\frac{\theta_4(0,x^3)}{\theta_4(0,x)}&=&
\prod_{k=1}^{\infty}\frac{(1+x^{k})}{(1-x^{k})}\frac{(1-x^{3k})}{(1+x^{3k})}\nonumber\\
&=&\prod_{k=1}^{\infty}\frac{(1-x^{3k})}{(1-x^{k})}
\frac{(1-x^{6k-3})}{(1-x^{(2k-1)})},
\end{eqnarray}
where we have used the Euler's theorem in obtaining the last
identity. Next changing $x$ to $x^2$, then we have,
\begin{eqnarray}
\label{toto46}
\frac{\theta_4(0,x^{2(3)})}{\theta_4(0,x^2)}&=&
\prod_{k=1}^{\infty}\frac{(1-x^{2(3k)})}{(1-x^{2k})}
\frac{(1-x^{2.3(2k-1)})}{(1-x^{2(2k-1)})}\nonumber\\ &=&
\prod_{k=1}^{\infty}\frac{(1-x^{3(2k)})}{(1-x^{2k})}
\frac{(1-x^{3.2(2k-1)})}{(1-x^{2(2k-1)})},
\end{eqnarray}
this gives us two definitions, the first one is that the partition
function $\frac{\theta_4(0,x^3)}{\theta_4(0,x)}$ is the generating
function for partitions of $2k$ prime to $3$ in which all odd parts
occur with even multiplicities. There is no restriction on the even
parts. This follows from the first identity in which both even and
odd parts are free of multiples of $3$. There is no restrictions on
the other even parts. There is an other definition that follows from
the second identity which gives the number of partitions of $2k$ in
which odd parts occur with multiplicity $2$ or $4$. The even
parts appear with at most twice. Using the second definition, then $6$ can be written as
\begin{center}
$6=4+2=4+1+1=2+2+1+1=3+3=2+1+1+1+1$,\end{center} and so $a(6)=6$.
The above two definitions can be  generalized for all $s$, this is given by  following proposition
\begin{proposition}
The generating function $\frac{\theta_4(0,x^s)}{\theta_4(0,x)}$, for
$s$ odd is the generating function for partitions of $2k$ into parts
prime to $s$. All odd parts occur with even multiplicities, there is
no restrictions on the even parts. In general for all $s$, it is a
generating function for partitions of $2k$ into parts in which all
odd parts occur with multiplicities $2,4, \cdots,2(s-1)$. The even
parts occur with at most $s-1$ times.
\end{proposition}
Proof: the proof is the same
as we did for $s=3$,
\begin{eqnarray}
\label{toto47}
\frac{\theta_4(0,x^s)}{\theta_4(0,x)}&=&
\prod_{k=1}^{\infty}\frac{(1+x^{k})}{(1-x^{k})}\frac{(1-x^{sk})}{(1+x^{sk})}\nonumber\\
&=&\prod_{k=1}^{\infty}\frac{(1-x^{sk})}{(1-x^{k})}
\frac{(1-x^{s(2k-1)})}{(1-x^{(2k-1)})}.
\end{eqnarray}
Next, change $x$ to $x^2$, then,
\begin{eqnarray}
\label{toto48}
\frac{\theta_4(0,x^{2(s)})}{\theta_4(0,x^2)}&=&
\prod_{k=1}^{\infty}\frac{(1-x^{2(sk)})}{(1-x^{2k})}
\frac{(1-x^{2.s(2k-1)})}{(1-x^{2(2k-1)})}\nonumber\\ &=&
\prod_{k=1}^{\infty}\frac{(1-x^{s(2k)})}{(1-x^{2k})}
\frac{(1-x^{s.2(2k-1)})}{(1-x^{2(2k-1)})},
\end{eqnarray}
so, for $s$  odd both the unrestricted even parts
$1/(1-x^{2k})$ and the odd parts that occur with even multiplicity
$1/(1-x^{2(2k-1)})$ are prime to $s$. From the last identity, we
learn that for $s$ even or odd, the even part occur with at most
$s-1$ while the odd parts  occur with multiplicities
$2,4,\cdots,2(s-1)$.

 A sequence A080054
\cite{sloane}, whose generating function is
\begin{equation}
\label{toto49}
 \frac{\theta_4(0,x^2)}{\theta_4(0,x)}=
\prod_{k=1}^{\infty}\frac{(1+x^{2k-1})}{(1-x^{2k-1})}=\prod_{k=1}^{\infty}(1+x^{2k-1})(1+x^k)
\end{equation}
corresponds the number of partitions of $2k$ in
which the odd parts occur with multiplicity $2$. The even parts
occur with multiplicity one. This definition was given by the author some time ago, to see it change  $ x  $ to $ x^2$ in the expression of the last product of equation (\ref{toto49}). Setting $a=-1$ in the
Lebesgue's identity \cite{andrews}, corollary (2.7), then
\begin{equation}
\label{toto50}
\frac{\theta_4(0,x^2)}{\theta_4(0,x)}=
1+\sum_{k=1}^\infty\frac{2(1+x)(1+x^{2})\cdots(1+x^{k-1})}
{(1-x)(1-x^{2})\cdots(1-x^{k})}x^{k(k+1)/2}.
\end{equation}

It is interesting to note that this partition function together with the partition function of a parafermion of order $4$,  which may be written as,  $ Z_{4}(\beta)=\prod_{k=1}^{\infty}\frac{(1+x^{2k})}{(1-x^{2k-1})}$  are nothing but the Ramond characters of the superconformal models  $SM(2,8)$ \cite { Fortin2}, and in general the partition function $ \frac{\theta_4(0,x^k)}{\theta_4(0,x)}$ for all $k$ is identified with  the Ramond character  $ \hat{\chi}_{1,2k}^{(2,4k)}(q)$ of $SM(2,4k)$. 
While writing this paper, we realized what there is a connection
between our work and that of overpartitions used by Corteel and
Lovejoy in combinatorial proofs of many q-series identities
\cite{corteel}. There is also an alternative definition that arises
from conformal field theory called jagged partitions \cite{Fortin1}. An overpartition of $k$ is an ordered sequence of nonincreasing
integers whose sum is $k$, where the first occurrence of each
distinct integer (equivalently, the final occurrence) is overlined. Therefore,
distinct parts are overlined and the unrestricted parts are non-overlined. For example the $8$ overpatitions of $3$ are \begin{center}$ 3,
\bar{3}, 2+1, \bar{2}+1, {2}+\bar{1}, \bar{2}+\bar{1}, \bar{1}+1+1,
1+1+1.$ \end{center} This is exactly the number of partitions of $6$
that we considered above.   Note that this definition coincides with the tensor
product of the bosonic and fermionic Fock spaces discussed above. If
we denote $\bar{p}(k)$ the number of overpartitions, then

\begin{eqnarray}
\label{toto51}
\frac{Z_{s}(\beta)}{\Delta_s^{+}(\beta)} &=&
\prod_{k=1}^{\infty}\frac{(1+x^{k})}{(1-x^{k})}\nonumber\\ &=&
\sum_{k=0}^\infty\bar{p}(k)x^k \nonumber\\
&=&1+2x+4x^{2}+8x^{3}+14x^{4}+24x^{5}+\cdots.
\end{eqnarray}
The other interesting example, is when $ s=3$ for which the number of partitions of $6$ prime to $3$ is $6$.
In the language of overpartitions, we are dealing
with overpartitions in which the parts are not divisible by $3$.
Therefore, discarding $3$ and $\bar{3}$ then the number of
overpartitions of $6$ is $6$. This also coincides with our second
definition of $\frac{\theta_4(0,x^3)}{\theta_4(0,x)}$.  The
correspondence between overpartitions in which the parts are not
divisible by $s$, and our definitions for partitions of $2k$ prime to
$s$ is as follows; First,  if $s$ is odd, the number of
overpartitions not divisible by $s$ is the number of partitions of
$2k$ prime to $s$. All odd parts occur with even multiplicities,
there is no restrictions on the even parts. Second , if  $s$ is even,
then, the number of overpartitions not divisible by $s$ is the number
of partitions of $2k$ in which all odd parts occur with
multiplicities $2,4, \cdots,2(s-1)$. The even parts appear with at
most $s-1$ times. In particular, if  $s=2$, one has
\begin{eqnarray}
\label{toto52}
\frac{\theta_4(0,x^2)}{\theta_4(0,x)}&=&
\prod_{k=1}^{\infty}\frac{(1+x^{k})}{(1-x^{k})}\frac{(1-x^{2k})}{(1+x^{2k})}\nonumber\\
&=& 1+2x+2x^2+4x^3+6x^4+8x^5+\cdots,
\end{eqnarray}
this tells us that the number of overpartitions of $3$ in which no
part is divisible by $2$ is $4$. we can see this by going back to
the above example, then discarding partitions that involve $2$ and
$\bar{2}$.
Using Andrews's multiple series transformation, and the
Jacobi triple product \cite{andrews}, we can express
$\frac{\theta_4(0,x^s)}{\theta_4(0,x)}$ as a multiple sum
\cite{corteel},
\begin{equation}
\label{toto53}
\frac{\theta_4(0,x^s)}{\theta_4(0,x)}=
\sum_{n_{s-1}\geq \cdots \geq n_{1}\geq{0}}
\frac{\prod_{j=0}^{{n_{s-1}-1}}(1+x^j)x^{{1/2}{{n_{s-1}(n_{s-1}+1)}+n_{s-2}^{2}+\cdots
n_{1}^{2}}}}{\prod_{j=0}^{n_{s-1}-n_{s-2}-1}(1-x^{j+1})\cdots
\prod_{j=0}^{n_{2}-n_{1}-1}(1-x^{j+1})
\prod_{j=0}^{n_{1}-1}(1-x^{j+1})}.
\end{equation}
The case $s=2$ follows from the Lebesgue's identity \cite{andrews}. Because of the connection
with the overpartitions, we conclude that our system whose partition
function is $\frac{Z_{s}(\beta)}{\Delta_s^{\pm}(\beta)}$, is given
by the single sum through the Lebesgue's identity for $s$ even.
However, for $s$  odd, the partition function is given by the
multiple sum. We finish this section by noting that if the order
of parafermions is even, then the system with the  partition function  (s-IGPPF) may be
described in terms of overpartitions. This is clear from the
expression of the partition function $\frac{1}{\Delta_s^{+}(\beta)}$ computed in the
last section, see equation (\ref{toto35}), then
\begin{eqnarray}
\label{toto54}
Z_F(\beta)\Delta_B(s\beta)&=&\prod_{k=1}^{\infty}\frac{(1+x^{k})}
{(1-x^{sk})}\nonumber\\
&=&\prod_{k=1,k\nmid{s/2}}^{\infty}(1+x^{k})\prod_{k=1}^{\infty}\frac{1}
{(1-x^{(s/2)k})},
\end{eqnarray}
this partition function corresponds to a tensor (convolution)
product of a certain bosonic Fock space with a fermionic Fock space,
and has a similar form of overpartitions but with restrictions. From
the above identity we have two definitions associated to this type
of overpartitions. The first definition is like the above definition
of  overpartition of $k$ for the overlined parts. However, the
non-overlined parts occur with multiplicity $s$. The second
definition, implies that distinct parts which are overlined are not
divisible by $s/2$, and the non-overlined (unrestricted) parts are
multiples of $s/2$. If $s=4$, we know that the number of partition
of $7$ is $7$, therefore in terms of overpartitions, the $7$
overpatitions of $7$ are given by
\begin{center}$ \bar{7}, \bar{6}+\bar{1}, \bar{5}+\bar{2},\bar{4}+\bar{3},
4+\bar{3}, \bar{4}+\bar{2}+\bar{1},
4+\bar{2}+\bar{1}.$\end{center} or
\begin{center}$ \bar{7}, 6+\bar{1}, \bar{5}+2, 4+\bar{3},
4+2+\bar{1}, \bar{3}+2+2, 2+2+2+\bar{1},$\end{center} using the
second definition.

\section{CONNECTION WITH THE RIEMANN GAS}
The Riemann gas is a system whose partition function is given by the
Riemann zeta function \cite{Julia},
\begin{eqnarray}
\label{toto55}
\zeta(t)&=&\frac{1}{\prod_{p}(1-p^{-t})}\nonumber\\
&=&\sum_{n=1}^{\infty}1/n^{t},
\end{eqnarray}
where the product is over all prime numbers $p$ and
$t=\beta\omega$. The different partition functions obtained so far
are related to the generating functions of partitions of integers.
Therefore we may call these theories additive quantum theories to
differentiate them from multiplicative quantum theories whose
partition functions are related to the zeta function. Using the
fact that bosonic partition functions for the Riemann gas and the
Euler gas are given by infinite products. Therefore we may go from
additive quantum theory to multiplicative quantum theory, simply
by changing product over integers into product over primes, and
changing the $x^k$ by $p^{-t}$. For example the fermionic
partition function in the additive quantum theory is
$Z_{F}(\beta)=\prod_{k=1}^{\infty}(1+x^{k})$ which becomes in the
multiplicative theory as
\begin{eqnarray}
\label{toto56}
Z_{F}(t)&=&\prod_{p}(1+p^{-t})\nonumber\\
&=&\frac{\prod_{p}(1-p^{-2t})}{\prod_{p}(1-p^{-t})}\nonumber\\
&=&\frac{\zeta(t)}{\zeta(2t)}\nonumber\\
&=&\sum_{n=1}^{\infty}q(n)/n^{t},
\end{eqnarray}
where $q(n)=1$, if $n$ is a square free, and $q(n)=0$ if $n$ has a squared factor. Therefore, in the Fock space of states, if we label
the fermionic states
$\prod_{i}(f_{i}^{\dag})^{\beta_{i}}|\tilde{\emptyset}\rangle$ by
integers, that is, $|n\rangle$, $n=\prod_{i}p_{i}^{\beta_{i}}$ then
$\beta_{i}$ must take the values $0$ or $1$. Note that $q(n)$ is
related to the M\"{o}bius function $\mu(n)$,  by $q(n)=|\mu(n)|$
\cite{hardy} where $\mu(1)=1$, $\mu(n)=(-1)^k$ if $n$ is the product
of different primes,  and $\mu(n)=0$, if $\beta_{i}>1$. Similarly, for
parafermions, using equation (\ref{toto16}), the parafermionic
partition function for the Riemann gas can be written as,
\begin{eqnarray}
\label{toto57}
Z_{s}(t)&=&\prod_{p}(1+p^{-t}+p^{-2t}+\cdots+p^{-(s-1)t})\nonumber\\
&=&\frac{\prod_{p}(1-p^{-st})}{\prod_{p}(1-p^{-t})} \nonumber\\
&=&\frac{\zeta(t)}{\zeta(st)}\nonumber\\
&=&\sum_{n=1}^{\infty}q_{s}(n)/n^{t},
\end{eqnarray}
where $q_{s}(n)$ takes the values $0$ or $1$, depending if $n$ has an
$sth$ power as a factor or not. All of these partitions functions
are well known for the multiplicative quantum theory \cite{Spector}.
Next, we  will obtain the partition
functions that are  analogous  to those in the additive quantum theory. From
equation (\ref{toto23}), the graded partition function for $s$ even
is
\begin{eqnarray}
\label{toto58}
\Delta_s^{+}(t)&=&\prod_{p}(1-p^{-t}+p^{-2t}-\cdots-p^{-(s-1)t})\nonumber\\
&=&\frac{\prod_{p}(1-p^{-st})}{\prod_{p}(1+p^{-t})}\nonumber\\
&=&\frac{\zeta(2t)}{\zeta(t)\zeta(st)}\nonumber\\
&=&\sum_{n=1}^{\infty}\mu_{s}(n)/n^{t},
\end{eqnarray}
where we have introduced the truncated M\"{o}bius function
$\mu_{s}(n)$ function to differentiate it from the ordinary
M\"{o}bius function $\mu(n)$. The truncated M\"{o}bius  function
$\mu_{s}(n)$ would be zero if $n$ is divisible by the $sth$ power of
some prime, so we may call it the M\"{o}bius function of order $s$.
For example if $s=2$, we obtain;
$$\frac{1}{\zeta(t)}=\sum_{n=1}^{\infty}\mu(n)/n^{t},$$ that is, the usual
expression for the inverted zeta function. For $s$ odd, then, the
graded partition function is
\begin{eqnarray}
\label{toto59}
\Delta_s^{-}(t)&=&\prod_{p}(1-p^{-t}+p^{-2t}-\cdots +p^{-(s-1)t})\nonumber\\
&=&\frac{\prod_{p}(1+p^{-st})}{\prod_{p}(1+p^{-t})} \nonumber\\
&=&\frac{\zeta(2t)\zeta(st)}{\zeta(t)\zeta(2st)}\nonumber\\
&=&\sum_{n=1}^{\infty}\mu_{s}(n)/n^{t}
\end{eqnarray}
From these two equations, the counter parts of the inverted graded
partition functions (\ref{toto26}), (\ref{toto27}) follows. Thus, if $s$ is even, the inverted parafermionic partition function may be written as
\begin{eqnarray}
\label{toto60}
\frac{1}{\Delta_s^{+}(t)}&=&
Z_{B}(st)Z_{F}(t)\nonumber\\
&=&\frac{\prod_{p}(1+p^{-t})}{\prod_{p}(1-p^{-st})} \nonumber\\
&=&\frac{\zeta(t)\zeta(st)}{\zeta(2t)} ,\nonumber\\
&=&\prod_{p}(\sum_{j=0}^{\infty}c(j)^{+}p^{-tj}).
\end{eqnarray}
Where $c(0)^{+}=1$, $c(j)^{+}=1$ if $j$ is congruent to $ 0,1$
modulo $s$ and $c(j)^{+}=0$ if $j$ is congruent to
$2,3,\cdots,(s-1)$ modulo $s$. Therefore, if we write the
partition function $\frac{\zeta(t)\zeta(st)}{\zeta(2t)}$
additively as the Dirichlet series
$\sum_{n=1}^{\infty}a(n)^{+}n^{-t}$, then the coefficients
$a_{n}^{+}$ are multiplicative with $a^{+}(p_{1}^{j_1}\cdots
p_{r}^{j_{r}})= c(j_{1})^{+}\cdots c(j_{r})^{+}$. In particular
$a(n)^{+}=0$ if $n$ is divisible by some prime to an exponent
which is congruent to $2,3,\cdots,(s-1)$ modulo $s$, and is equal
to $1$ otherwise. As a consequence, the Dirichlet formula for any $s$,
may be written as,
\begin{equation}
\label{toto61} \frac{\zeta(t)\zeta(st)}{\zeta(2t)}
=\sum_{n=1}^{\infty}\frac{1}{n^{t}},
\end{equation}
where the exponents in $n=\prod_{i}p_{i}^{r_{i}}$, $r_{i}$, are
congruent to $ 0,1$ modulo $s$. In terms of the quantum states
$|n\rangle$, in the Fock space labeled by an integer $n$,
$n=\prod_{i}p_{i}^{r_{i}}$, the states for which the exponents
${r_{i}}$ are congruent to $2,3,\cdots,(s-1)$ modulo $s$, are
missing. The only states present in the Fock space are those for
which the exponents are congruent to $ 0,1$ modulo $s$. If the  the
order of parafermions $s$ is odd, then  the
inverted parafermionic partition function for the Riemann gas would be;
\begin{eqnarray}
\label{toto62}
\frac{1}{\Delta_s^{-}(t)}&=&
Z_{F}(t)\Delta_{B}(st)\nonumber\\
&=&\frac{\prod_{p}(1+p^{-t})}{\prod_{p}(1+p^{-st})} \nonumber\\
&=&\frac{\zeta(t)\zeta(2st)}{\zeta(2t)\zeta(st)}\nonumber\\
&=&\prod_{p}(\sum_{j=0}^{\infty}c(j)^{-}p^{-tj}),
\end{eqnarray}
where $c(0)^{-}=1$, $c(j)^{-}=1$ if $j$ is congruent to $ 0,1$
modulo $2s$, $c(j)^{-}=-1$ if $j$ is congruent to $ 0,1$ modulo
$s$. Alternatively we may write, $c(j)^{-}=(-1)^{[j/s]}$, where
$[x]$ denotes the integer part of $x$, and $c(j)^{-}=0$ if $j$ is
congruent to $2,3,\cdots,(s-1)$ modulo $s$. In this case the
Dirichlet series associated with the partition function
$\frac{\zeta(t)\zeta(2st)}{\zeta(2t)\zeta(st)}$, is
$\sum_{n=1}^{\infty}a_{n}^{-}n^{-t}$, where the coefficients
$a_{n}^{-}=0$, if $n$ is divisible by some prime to an exponent
which is congruent to $2,3,\cdots,(s-1)$ modulo $s$. It is equal
to $\pm1$ otherwise. Note that the minus one is due to the graded
bosonic partition function $\Delta_{B}(st)$. Therefore, the
Dirichlet series takes the following expression;
\begin{equation}
\label{toto63}
\frac{\zeta(t)\zeta(2st)}{\zeta(2t)\zeta(st)}=
\sum_{n=1}^{\infty}\frac{(-1)^{[r_1/s]+[r_2/s]+\cdots+[r_k/s]}
}{n^{t}},
\end{equation}
where the exponents $r_{i}$ in $n=\prod_{i}p_{i}^{r_{i}}$, are not
congruent to $2,3,\cdots,(s-1)$ modulo $s$. In this case, the Fock
space of states labeled by an integer $n$,
$n=\prod_{i}p_{i}^{r_{i}}$, in which the exponents ${r_{i}}$ are
congruent to $2,3,\cdots,(s-1)$ modulo $s$ are missing, otherwise,  the states would be present. We give finally, the analogue of the generating functions $\frac{1}{\theta_4(0,x)}$ and $
\frac{\theta_4(0,x^s)}{\theta_4(0,x)}$, given by equations
(\ref{toto33}), (\ref{toto34}) respectively, for the Riemann gas.
Following the same procedure as above, the analogue of
$\frac{1}{\theta_4(0,x)}$ is;
\begin{eqnarray}
\label{toto64}
\frac{Z_{s}(t)}{\Delta_s^{+}(t)}&=&
Z_{B}(t)Z_{F}(t),\nonumber\\
&=&\frac{\prod_{p}(1+p^{-t})}{\prod_{p}(1-p^{-t})} \nonumber\\
&=&\frac{\zeta(t)^2}{\zeta(2t)} ,\nonumber\\
&=&\prod_{p}(1+2p^{-t}+2p^{-2t}+2p^{-3t}+\cdots),\nonumber\\
&=&\prod_{p}(\sum_{j=0}^{\infty}2^{\nu(p^j)}p^{-tj}),\nonumber\\
&=&\sum_{n=1}^{\infty}2^{\nu(n)}/n^{t},
\end{eqnarray}
where the sum is over all integers $n$, and $\nu(n)$ is the number
of distinct prime factors of $n$, and so $\nu(1)=0$ and if
$n=\prod_{i=1}^{k}p_{i}^{r_{i}}$, then $\nu(n)=k$. This is exacly the  partition  function associated with the unitary twisting of two identical bosonic
gases  \cite{i.bakas}. Here, we obtain this result by mixing a paraferminic system
with a system whose partition function is the inverted graded
partition function. As a result, the analogue of
$\frac{1}{\theta_4(0,x)}$ in the Riemann gas is $\frac{\zeta(t)^2}{\zeta(2t)}$. We may derive  equation (\ref{toto64}), by computing the
total partition function $Z_T(\beta)= {\rm Tr}\exp{-\beta H}$, where $H$ is the total Hamiltonian, i.e, $H=H_B+H_F$, and the
trace is taken over special kind of normalized states that we
denote by $|n,d \rangle $, here, $n$ is the total number
eigenvalue such that the numbers $d$ are divisors of $n$, and  square free \cite{donald}. The state $|n,d\rangle$, in the multiplicative
quantum theory has an energy $E_n=\beta\log(n)$. Writing
$E_{n}=\langle n,d|H|n,d\rangle$, and using the notation
${\sum}'_{d\mid n}$ to mean that the sum is over all divisors $d$ of $n$
which are square free, then the total partition function becomes,
\begin{eqnarray}
\label{toto65}
Z_T(\beta)&=&\sum_{n=1}^{\infty}({\sum}'_{d\mid n}
\langle n,d
|\exp{-\beta H}|n,d\rangle)\nonumber\\
&=&\sum_{n=1}^{\infty}\exp{-\beta
E_n}({\sum}'_{d\mid n} \langle n,d|n,d\rangle),\nonumber\\
&=&\sum_{n=1}^{\infty}\exp-{(\beta \omega\log(n))}\sum_{d\mid
n}q(d),
\end{eqnarray}
where $q(d)= \langle n,d|n,d\rangle$ is equal to $1$, when $d$ is
a square free, and is equal to $0$, when $d$ has squared factor.
Note that if we did not have the constraint that $d$ is a square
free but just a divisor of $n$, then $$ \sum_{d\mid n}\langle
n,d|n,d\rangle=\sum_{d\mid n}1=\tau(n)$$ that is, the number of
divisors of $n$. However, in our case the sum is given by the
number of unitary divisors of $n$ through the identity
$\sum_{d\mid n}q(d)=2^{\nu(n)}$, recall that $d$ is a unitary
divisor of $n$ if the greatest common divisor of $d$ and
$\frac{n}{d}$ is one. Using the latter identity, and setting
$t=\beta\omega$ in the above equation (\ref{toto65}), gives
\begin{equation}
\label{toto66}
Z_T(\beta)=\sum_{n=1}^{\infty}2^{\nu(n)}/n^{t}.
\end{equation}
This is exactly equation (\ref{toto64}), this would be, the  the analogue of the  overpartition generating function in the additive
number theory. As an example, the bosonic and Fermionic states in
the Fock space associated to the mixed system, whose partition
function is given by equation (\ref{toto64}), up $n=20$ are,
\begin{center}$|1\rangle|, |2\rangle, |3\rangle, |4\rangle,
|5\rangle, \cdots, |20\rangle, $\end{center} for the bosonic states,
and
\begin{center}$|1\rangle|, |2\rangle, |3\rangle, |5\rangle,
|6\rangle, \cdots, |19\rangle, $\end{center} for the fermionic
states, note that in this case, the integers are square free. One
can see that the tensor product of these states defined by
multiplying the bosonic labels times the fermionic labels in this
order, such that the total number $n=20$, would corresponds to the
following states,
\begin{center}$|20.1\rangle, |10.2\rangle, |4.5\rangle,
|2.10\rangle. $\end{center}
 Using the notation $|n,d \rangle $,  where $d$ is a square free divisor of $20$, then, the  $4$ states are;
\begin{center}$|20,10\rangle, |20,5\rangle, |20,2\rangle,
|20,1\rangle, $\end{center} alternatively in terms of unitary
divisors $d^{*}(20)$, this is written as,
\begin{center}$|20,20\rangle, |20,5\rangle, |20,4\rangle,
|20,1\rangle, $\end{center} this simply says, that the number of
square free divisors of $n$ is the same as the number of unitary
divisors $d^{*}(n)$ of $n$. Therefore, we may alternatively denote
the states $|n,d\rangle$, by $|n, d^{*}(n)\rangle$. Alternatively, we may relate the additive and multiplicative theories  through  the correspondence between the additive generating theta function $ \theta(t)$ and the multiplicative Riemann zeta function $\zeta(t) $. This correspondence states that the the Riemann zeta function is the  Mellin transform  of $( \theta(t)-1 )$ \cite{Koblitz}. The Mellin transform of a given function $f(t)$ on the positive real line  is defined by the formula  $ \Phi(s):=\int_{0}^{\infty}f(t)t^{s-1}dt$ for values of $s$ such that the integral converges. We will show  up to some factors that  the Mellin transform of the theta Jacobi $ \theta_{4}(t)$ is  the alternating Riemann zeta function, also called the Dirichlet Eta function $\eta(s)=\sum_{n=1}^\infty (-1)^{n+1}\frac{1}{n^{s}}$. This result was given in \cite{Glasser}, however, our derivation is explicit and  closely related to  the derivation that makes  the connection between $ \theta(t)$ and  $\zeta(t) $ functions. To this  end, let us first write $ \theta_{4}(t)= \sum_{n=-\infty}^\infty (-1)^n e{ ^{-n^{2} \pi t}}$ in terms of $\theta(t)=1+2 \sum_{n=0}^\infty  e{ ^{-n^{2} \pi t}} $. The theta function $ \theta_{4}(t)$ may be written explicitly in the form
\begin{eqnarray}
\label{theta1}
\theta_{4}(t)&=&1+2\big(\sum_{n\ even}^\infty  e{ ^{-n^{2} \pi t}}-\sum_{n\ odd}^\infty  e{ ^{-n^{2} \pi t}}\big)\nonumber\\&=&1+2\big(2\sum_{n=1}^\infty  e{ ^{-4n^{2} \pi t}}-\sum_{n=1}^\infty  e{ ^{-n^{2} \pi t}}\big)\nonumber\\&=&2\theta(4t)-\theta(t)
\end{eqnarray}
To find  the Mellin transform of $\theta_{4}(t) $, we  use the fact that  the theta function $ \theta(t)$ satisfy the following functional equation
\begin{equation}
\label{theta2}
\theta(t)=t^{-1/2}\theta(1/t),
\end{equation}
and so, 
\begin{equation}
\label{theta3}
\theta(4t)=\frac{1}{2}t^{-1/2}\theta(1/4t).
\end{equation}
Thus, near $t=0$, $\theta(t)$ looks like  like $ t^{-1/2}$, and  $\theta(4t)$,   looks like $\frac{1}{2}t^{-1/2}$, while for large $t$ both $\theta(t)$ and $\theta(4t)$ are asymptotic to $1$.  Usually In obtaining the zeta function $\zeta(t)$ from the theta function, we replace  $s$ by $s/2$ in the definition of the Miller transform. Therefore,the Mellin transform may be written as 
\begin{equation}
\label{theta4}
\Phi(s)(\theta(4t)):=\int_{1}^{\infty}t^{s/2-1}(\theta(4t)-1)dt)+\int_{0}^{1}t^{s/2-1}(\theta(4t)-\frac{1}{2}t^{-1/2})dt).
\end{equation}
Note that  the first integral is convergent  for large $t$ , while the second integral is convergent near $t=0$. Now, it is safe to integrate term by term in the above integrals and at the end we sum over $n$ to obtain
\begin{eqnarray}
\label{theta5}
\Phi(s)(\theta(4t)):&=&\int_{1}^{\infty}t^{s/2-1}(\theta(4t)-1)dt)+\int_{0}^{1}t^{s/2-1}(\theta(4t)-\frac{1}{2}t^{-1/2})dt)\nonumber\\&=&\frac{2}{\pi^{s/2}2^s}\Gamma(s/2)\zeta(s)+\frac{2}{s}+ \frac{1}{1-s}.
\end{eqnarray}
Now, the Mellin transform of $\theta(t)$ is given by the formula
\begin{eqnarray}
\label{theta6}
\Phi(s)(\theta(t)):&=&\int_{1}^{\infty}t^{s/2-1}(\theta(t)-1)dt)+\int_{0}^{1}t^{s/2-1}(\theta(t)-t^{-1/2})dt)\nonumber\\&=&\frac{2}{\pi^{s/2}}\Gamma(s/2)\zeta(s)+\frac{2}{s}+ \frac{2}{1-s} .
\end{eqnarray}
This is the correspondence between the additive theta function and the multiplicative  Riemann zeta function.
Finally, the Mellin transform of the theta function $ \theta_{4}(t)$ may be given by the following formula
 \begin{eqnarray}
\label{theta7}
\Phi(s)(\theta_{4}(t)) &=& 2\Phi(s)(\theta(4t))-\Phi(s)(\theta(t))\nonumber\\&=&\frac{2}{\pi^{s/2}}\Gamma(s/2)\big(2^{1-s}-1)\zeta(s)\big) +\frac{2}{s}\nonumber\\&=& -\frac{2}{\pi^{s/2}}\Gamma(s/2)\eta(s)+\frac{2}{s} .
\end{eqnarray}
Therfore, the altrnating theta function $ \theta_4(0,x)$ is related to the alternating Riemann zeta function $ \eta(s)$. Because of the Dirichlet series  $\sum_{n=1}^{\infty}2^{\nu(n)}/n^{t}$  representation of $\frac{\zeta(t)^2}{\zeta(2t)} $, then, one would  follow formally,  similar steps as in \cite{Serre} page $167$. Given the Drichlet series $\Psi_{f}(s)=\sum_{n=1}^{\infty}\frac{c(n)}{n^{s}}$, where $ f(z)=\sum_{n=0}^{\infty}e^{2\pi iz}$ is a modular form of weight $2k$, $k>0$,  Hecke proved that the series $\Psi_{f}(s)$  can be analytically extended to a meromorphic function in the whole complex pane, and the function defined by $ \Lambda_{f}(s)=(2\pi)^{-s}\Gamma(s)\Psi_{f}(s)$  satisfies : $$\Lambda_{f}(s)=(-1)^{k}\Lambda_{f}(2k-s) .$$ F. Diamonde informed me that from the Euler product of the Dirichlet series, see equation (\ref{toto64}), the Euler factors are not the sort that could arise from a modular form. therefore, one should look for a non direct method, and can not be done as in the correspondence between the Riemann zeta function and the theta function.

 Finally, we would like to find the analogue of the partition function $
\frac{\theta_4(0,x^s)}{\theta_4(0,x)}$, first the theta for the Riemann gas, may be written for $s$ odd, as follows;
\begin{eqnarray}
\label{toto67}
\frac{Z_{s}(t)}{\Delta_s^{-}(t)}&=&
Z_{B}(t)Z_{F}(t)\Delta_{B}(st)\Delta_{F}(st),\nonumber\\
&=&\frac{\prod_{p}(1+p^{-t})}{\prod_{p}(1-p^{-t})}
\frac{\prod_{p}(1-p^{-st})}{\prod_{p}(1+p^{-st})}\nonumber\\
&=&\frac{\zeta(t)^2}{\zeta(2t)}\frac{\zeta(2st)}{\zeta(st)^2} ,\nonumber\\
&=&\prod_{p}\frac{(1+2p^{-t}+2p^{-2t}+2p^{-3t}+\cdots+2p^{-(s-1)t}+p^{-st})}{(1+p^{-st})},\nonumber\\
&=&\prod_{p}(\sum_{j=0}^{\infty}c(j)p^{-tj}),
\end{eqnarray}
where $c(0)=1$, $c(j)=0$ if $j>0$ and $j$ is divisible by $s$, and
$c(j)=2(-1)^{[j/s]}$ if $j$ is not divisible by $s$ , where $[x]$
denotes the integer part of $x$. Therefore, if we write the partition function
$\frac{\zeta(t)^2}{\zeta(2t)}\frac{\zeta(2st)}{\zeta(st)^2}$,
additively as the Dirichlet series $\sum_{n=1}^{\infty}a(n)n^{-t}$.
Then the  multiplicative function $a(n)$ takes the following values,
$a(n)=0$ if $n$ is divisible by some prime to an exponent which is
a multiple  $s$. Otherwise, it is equal to $\pm1$ times power of 2 .
Therefore, the quantum states of the Riemann gas for $s$
odd, labeled by an integer $n$, where $n=\prod_{i}p_{i}^{r_{i}}$ are
free of primes whose powers are divisible by $s$. This is similar to
the Euler gas whose partition function is $
\frac{\theta_4(0,x^s)}{\theta_4(0,x)}$, with $s$ odd. Recall, that in
this case, the subscripts of the bosonic and fermionic creation
operators are prime to $s$. By doing explicit computation for
different values of $s$, it turns out, that a general formula for
$\frac{\zeta(t)^2}{\zeta(2t)}\frac{\zeta(2st)}{\zeta(st)^2}$ may be
written as;
\begin{eqnarray}
\label{toto68}
\frac{\zeta(t)^2}{\zeta(2t)}\frac{\zeta(2st)}{\zeta(st)^2}
&=&\prod_{p}\frac{(1+2p^{-t}+2p^{-2t}+2p^{-3t}+\cdots+2p^{-(s-1)t}+p^{-st})}{(1+p^{-st})},\nonumber\\
&=&\sum_{n=1}^{\infty}\frac{(-1)^{[k_1/s]+[k_2/s]+\cdots+[k_r/s]}
2^{\nu(n)}}{n^{t}},
\end{eqnarray}
where the  sum is taken over all $n=p_{1}^{k_{1}}\cdots
p_{r}^{k_{r}}$, such that the exponents $k_{i}$'s are not divisible
by $s$. 
\section{DISCUSSION}
In this paper, we derived the Euler
theorem as well as a theorem in the theory of
partitions which may be stated as follows;  the number of partitions of $k$ in
which no parts appear more than $s-1$ times equals the number of
partitions of a positive integer $k$ such that no parts is divisible by $s$. The derivation  of the Euler theorem and its generalization is based on the fact that  the bosonic Hamiltonian of a non-interacting quantum field theory  may be decomposed into  bosonic Hamiltonian whose frequencies are twice ($s$-times) as much as the original Hamiltonian, and a fermionic (prafermionic) Hamiltonian.  By
realizing that the graded fermionic partition function
$\Delta_F(\beta)=\prod_{k=1}^{\infty}(1-x^{k})$ is obtained from
fermionic partition function by simply changing $x^k$ to $-x^k$. Then the expressions for the graded parafermionic partition
functions were obtained whose inverse would correspond to to
bosonic(fermionic) partition function depending on the  order of
parafermions being  even or odd, respectively. Both of these partition
functions, generate partition of numbers with given restrictions.
The generating functions  that we have obtained are general in the sense
that when the order of the  parafermions is two, then our
generating function coincides with the Euler generating
function. These generating functions give rise to some sequences of
partitions \cite{sloane}. If the order of the parafermions is
infinite, then the inverse of the  parafermionic partition functions ($s$-IGPPF) $\frac{1}{\Delta_s^{\pm}(\beta)}$ tend to a  fermionic partition
function. Therefore, as the boson is an infinite order
parafermion, a fermion would be an infinite order inverted graded
parafermion. The Euler theorem equates the number of distinct
partitions with the number of unrestricted odd partitions.The proof of this theorem is simple in number
theory,  this is also the case through the factorization of  the fermionic partition function into a bosonic partition function and a graded fermionic partition function. One may look
at it differently, and write
$Z_f(\beta)=\Delta_{F}(2\beta)/\Delta_{F}(\beta)$ from which 
the graded fermionic partition function is  $\Delta_{F}(2\beta)=Z_f(\beta)\Delta_{F}(\beta)$,  so
mixing the fermionic system with the graded fermionic system at
thermal equilibrum at a given  temperature $\beta$ is the same as a graded
parafermionic system whose  temperature is doubled. This also happens
in the case of quantum field theory with a logarithmic spectrum
\cite{Spector}, in which the term duality was used to characterize
the identities among arithmetic quantum theories. The above mixing
is a special case of our relation
$\frac{1}{\Delta_s^{+}(\beta)}=Z_B(s\beta)Z_F(\beta)$.
Also,  we may write the fermionic partition function in terms of
parafermionic partition functions as 
$Z_f(\beta)=\Delta_{F}(s\beta)/\Delta_s^{+}(\beta)$ or
equivalently
$Z_f(\beta)=\frac{1}{\Delta_{B}(s\beta)\Delta_s^{-}(\beta)}$. Mixing  a parafermion system of even order $s$  with a
system whose partition function is
$\frac{1}{\Delta_s^{+}(\beta)}$,  was shown to be equivalent
to mixing fermions and bosons. For $s$ odd, the system obtained is
still a mixing of fermions and bosons but with some states
missing in the Fock space of states. Also, in this case we related our partitions functions
with partition theory. We have  shown  that  these partition functions are written in terms of the
Jacobi theta function $\theta_4(0,x)$, in particular,  if $s$ is even, the partition function is $1/\theta_4(0,x)$, while if $s$ is odd then
the partition may be written as  $\theta_4(0,x^s)/\theta_4(0,x)$. Here, it is interesting  to note that the sequences generated by the partition functions  ($4$-IGPPF), and  $1/\theta_4(0,x)$ appear  in the counting of the Moore-Read edge spectra \cite{Michael}. On the other hand, the sequences associated  with the partition function of a parafermion of order $4$,  $ Z_{4}(\beta)$ with  partition function $\theta_4(0,x^2)/\theta_4(0,x)$  are those that appear in  the Ramond characters of the superconformal models  $SM(2,8)$ \cite { Fortin2}. In general the partition function $ \frac{\theta_4(0,x^s)}{\theta_4(0,x)}$ for all $s$ is identified with  the Ramond character  $ \hat{\chi}_{1,2s}^{(2,4s)}(x)$ of $SM(2,4s)$. 
 We have also given  a general definition of the  partition function$ \frac{\theta_4(0,x^s)}{\theta_4(0,x)}$ for 
all $s$ in terms of partition theory.  A connection
with overpartitions \cite{corteel} in which the parts are not
divisible by $s$ was also given. Using the expression for the parafermionic partition function given by Eq. (\ref{toto13}), one has
$\Delta_F(s\beta)=Z_s(\beta)\Delta_{F}(\beta)$ for any $s$ , this identity will be a trivial identity when $s$ goes to
infinity as both bosonic and graded fermionic partition functions
cancel each other, this is equivalent to  the Witten index which is a
complete cancelation between boson and fermions.
We have seen that the square of a bosonic operator is not bosonic operator but a parafermion with order  parastatistics   $p=1/2$, $p=3/2$. For a single oscillator, the norm of the states is positive in agreement with the expressions of the partition functions. For the quantum field theory associated with these oscillators, the norm is not always positive and so we expect to see this at the level of the partition function. In the case of a  parastatistical systems  obeying the Green's trilinear relations \cite{Green} with the total number of particles $N$, and with finite number of quantum states denoted by $M$. Then the  grand canonical partition function turns out to be   given in terms of the Schur's function  \cite{ Chatervedi}, $$s_{\lambda}(x_{1},x_{2},\cdots,x_{M})=\frac{det\big(x_{i}^{\lambda_{j}+M-j}\big)}{{det\big(x_{i}^{M-j}\big)}}; \\1\leq i,j\leq M,$$ where  $\lambda= (\lambda_{1},\lambda_{2},\cdots,\lambda_{M})$ is a partition of length  $l(\lambda) $ $\leq N$. Explicitly, the grand canonical partition function for parabose statistics of order $s$ is  $$ Z_{B}^{s}=\sum_{l(\lambda)\leq s}s_{\lambda}(x),$$ where the sum is over all partitions $\lambda$ such that $ l(\lambda) \leq s $. For  parafermi statistics, the grand canonical partition function may be written as $$Z_{F}^{s}=\sum_{\lambda_{1}\leq s}s_{\lambda}(x)=\frac{det\big(x_{i}^{s+2M-j}-x_{i}^{j-1}\big)}{{det\big(x_{i}^{2M-j}-x_{i}^{j-1}\big)}},  \\1\leq i,j\leq M,$$ 
here, $\lambda_{1}=l(\lambda\prime)$ and $\lambda\prime $ being the conjugate partition of the partition $\lambda$. If the order of the parastatistics is infinite then, there is no restrictions on the above sums and as a result \cite{Macdonald}, one obtains
$$ Z_{B}^{\infty}=Z_{F}^{\infty}= \prod_{i}\frac{1}{(1-x_{i})}\prod_{i< j}\frac{1}{(1-x_{i}x_{j})}.$$
This equality is a expected from the Green's trilinear relations in which  for $ s \rightarrow\infty$,  the parabose and the parafermi  statistics are identical \cite{Messiah}. The parabosonic partition function $ Z_{B}^{\infty}$ was derived earlier \cite{Hama2}, using the standad states of Ohnuki and Kamefuchi \cite{Ohnuki}, the above partition function was then used to obtain the expression for the partition function of the parabosonic string $Z$, as well as the large $n$ level density $d_{n }$ given by $Z=\sum d_{n} x^{n}$.  The parafermi partition function of order $s=3$ with two level system was computed in \cite {Hama3}. However, for $s>3$,  $M>s$, the expressions for the partition functions are lacking. 

The asymptotic partition of the parafermionc partition function $$ Z_{s}(\beta)=\prod_{k=1}^{\infty}\Big(\frac{1-x^{ks}}{1-x^{k}}\Big)= \sum_{n=0}^{\infty}p(n,s)x^{n}$$  was carried out by Hagis \cite{Hagis},  using the Hardy-Ramanujan-Rademacher method. And for $n \gg s$, his results are,
$$ p(n,s)\sim \frac{\sqrt12{} s^{1/4}}{(1+s)^{3/4}(24)^{3/4}}e^{\pi\sqrt{2s/[3(1+s)]}\sqrt{n}} .$$ For the parabosonic partition function discussed in this paper, that is, $$  \prod_{k=1}^{\infty}\Big(1+\frac{x^{ks}}{1-x^{k}}\Big)=\sum_{n=0}^{\infty}p(n,s)^{\prime}x^{n}.$$ One would like to give the  meaning  to $ p(n,s)^{\prime}$ in terms of partitions, as well as to consider the large $n$ limit for $p(n,s)^{\prime }$. One may follow the alternative derivation of Hagis formula, given in \cite{Miles} in which they obtained an expression for $\ln p(n,s)$ by mapping a mathematical problem to a physical problem. 
 \\ In the last section we obtained  the counter parts  of the  partitions functions associated with the Euler gas. This
was possible,  because both the Euler gas and the Riemann gas have a
bosonic partition function that are infinite products one is over
integers (the Euler partition function) and the other is over
primes(the Riemann zeta function). In particular, the counter part  of
$\frac{1}{\theta_4(0,x)}$ and $
\frac{\theta_4(0,x^s)}{\theta_4(0,x)}$, are 
$\frac{\zeta(t)^2}{\zeta(2t)}$,
$\frac{\zeta(t)^2}{\zeta(2t)}\frac{\zeta(2st)}{\zeta(st)^2}$, respectively. Thus,  we see   that the analogue  of the Jacobi Theta function is $\frac{\zeta(2t)}{\zeta(t)^2}$.  In
terms of the Fock space of states, for the Euler gas, the quantum states are constructed
from the operators $( a_k^{\dag})$, such that $k$ are prime to
$s$. For the Riemann gas, the quantum states $|n\rangle$ are those
states for which the integers $n$ are free of primes whose powers
are multiples of $s$. Also, note that all the the
partition functions obtained for the the Riemann gas are written
as a Dirichlet series which would be the analogue of the
Lebesgue-Cauchy and the Andrews's multiple sums \cite{corteel}. We
conclude this discussion, by giving other identities and make  connection between the
generating functions that come up in the  additive and the  multiplicative
number theory. From equations (\ref{toto58}) and (\ref{toto59}), we
have $$\sum_{n=1}^{\infty}\mu_{s}(n)/n^{t}=
\frac{\zeta(2t)}{\zeta(t)\zeta(st)}\;\mbox\;s\;\mbox{even}$$ and
$$\sum_{n=1}^{\infty}\mu_{s}(n)/n^{t}=
\frac{\zeta(2t)\zeta(st)}{\zeta(t)\zeta(2st)},\;\mbox\;s\;\mbox{odd}.$$
An explicit computation for $s$ odd in the first formula, and $s$
even in the second formula give the following identities,
\begin{equation}
\label{toto69}
\frac{\zeta(2t)}{\zeta(t)\zeta(st)}=
\sum_{n=1}^{\infty}2^{\nu_{s}(n)}\lambda(n)/n^{t},\;\mbox\;s\;\mbox{odd}
\end{equation}
\begin{equation}
\label{toto70}
\frac{\zeta(2t)\zeta(st)}{\zeta(t)\zeta(2st)}=
\sum_{n=1}^{\infty}2^{\nu_{s}(n)}\lambda(n)/n^{t},\;\mbox\;s\;\mbox{even}
\end{equation}
where we have introduced the function $\nu_{s}(n)$, which is equal
to $\nu(n)$, when the prime powers $r_{i}\geq s$, and if some
$r_{i}<s$, then $\nu_{s}(n)=k-\#(\hat{k_i})$, $\#(\hat{k_i})$ is
the number of prime whose powers are less then $s$. When $s=1$,
$\nu_{s}(n)=\nu(n)$ and we have the well known formula
\cite{apostol},
$$\frac{\zeta(2t)}{\zeta(t)^2}=
\sum_{n=1}^{\infty}2^{\nu(n)}\lambda(n)/n^{t}.$$  We saw that in the last
section that the generating function
\begin{equation}
\frac{\zeta(t)^2}{\zeta(2t)}=
\sum_{n=1}^{\infty}2^{\nu(n)}/n^{t},\nonumber\\
\end{equation}
is the counter part of the  generating function
$\prod_{k=1}^{\infty}\frac{(1+x^k)}{(1-x^k)}$. This is known to be
given by the Cauchy formula, see equation (\ref{toto39}), however,
to make the correspondence closer for the sum given by equation
(\ref{toto64}), we rewrite the latter product as,
$\prod_{k=1}^{\infty}(1+\frac{2x^k}{(1-x^k)})$. then an explicit
computation shows that the general expression for this product can
be written as 
\begin{equation}
\label{toto71}
\prod_{k=1}^{\infty}\frac{(1+x^k)}{(1-x^k)}=
1+\sum_{n\ge 1}\sum_{i\le n} \sum_{\begin{subarray}{l}n_1>n_2>\ldots>n_i>0\\
n_1+\ldots+n_i=n\end{subarray}}\frac{2^i x^n}
{(1-x^{n_1})\ldots(1-x^{n_i})}.
\end{equation}
Here,  $n_1>n_2>...>n_i$ are the part sizes of an overpartition, and 
$i$ be the number of part sizes (the number of distinct parts of $n$) and $n=n_1+...+n_i$ be the sum of the part
sizes. We have checked this formula for many values. It is
interesting to note that from the above formula, and its counter
part that the power of $2$, is the number of distinct parts for
all partitions of $n$ in the additive number theory. In the
multiplicative number theory, the power of $2$ is the number of
distinct primes factors of $n$. The above formula, shows clearly
that the expansion coefficients, are always even for $n\geq1$.  Also, the above formula suggests that One may relate an  overpatition  ${\tilde p(n)}$ with the ordinary partition $p(n)$, $2^{i}$ counts the number of distinct overpartitions, and $i$ is the number of distinct parts in the partition of $n$, so if $p_{i}(n)$ is the number of distinct parts in the partition of $n$, then from our  formula, one may write $${\tilde p(n)}=\sum_{i} 2^{i}p_{i}(n).$$ For example $p(4)=5 $, $4=3+1=2+2=2+1+1=1+1+1+1$, while $p_{1}(n)=3$,  $p_{2}(n)=2$, and so ${\tilde p(4)}=14$ in agreement with the overpartition of $4$. Slight modification of this formula may be used to count the number of overpations with restrictions. We know from the expression of overpartition modulo $2$, that  the the number of overpartions of $4$ in which no parts is divisible by $2$ is $6$. We can checke this using the above example by discarding the distinct parts divisible by $2$.  If ${\tilde p(n)^{2}}$, $p_{i}^{2}(n)$ denote the number overpartitions and the number of distinct parts modulo $2$, respectively, then, $p_{1}^{2}(4)=p_{2}^{2}(4)=1$, and so ${\tilde p(4)^{2}}=6$. Therefore, the general formula of the number of overpartitions such that no parts is divisible by $s$, would be $${\tilde p(n)^{s}}=\sum_{i}2^{i}p_{i}^{s}(n) .$$
Next, we propose a sum formula for the inverted graded parafermion
of order $4$,
$\prod_{k=1}^{\infty}\frac{(1+x^{2k-1})}{(1-x^{2k})}$. We learn
from Corteel that this generating was considered by Mac Mahon and
others, and the associated Ferrers diagram is called 2-modular
diagram \cite{berkovich}, \cite{pak}. The formula we propose is
given by following expression,
\begin{equation}
\label{toto72} \prod_{k=1}^{\infty}\frac{(1+x^{2k-1})}{(1-x^{2k})}
=1+\sum_{n=1,n\neq2}^{\infty}\sum_{m\geq{1}}\sum_{n_{1},\cdots
n_{m}}\frac{x^n (1+x)^{m}}
{(1-x^{n_1})(1-x^{n_2})\cdots(1-x^{n_m})},
\end{equation}
where $m$, is chosen such that $n+m$, is even and $n> m$ except
when $n=1$, $m=1$. The exponents $n_j$, are even and distinct
parts solutions to $n_1+n_2+\cdots+n_m=n+m$. Setting $s=4$, in
equation (\ref{toto61}), then the counter part of the above
formula in the multiplicative number theory, would be
\begin{equation}
\label{toto73}
\frac{\zeta(t)\zeta(4t)}{\zeta(2t)}
=\sum_{n=1}^{\infty}\frac{1}{n^{t}},
\end{equation}
where the exponents in $n=\prod_{i}p_{i}^{r_{i}}$, $r_{i}$, are
congruent to $ 0,1$ modulo $4$.\\
\vspace{7mm}

{\bf Acknowledgments:}
I would like to thank  S.Corteel for the combinatorial interpretation given to the formulas \ref{toto71} and \ref{toto72},  K.S.Narain, A.Lascoux, M.O'Loughlin, G.Thompson, for critical
reading of the manuscript and discussions, M.Somos, and D.Spector  for correspondence and D.Zagier for his kind suggestion on formula
(\ref{toto68}), F.Diamond for correspondence on modular forms, and the Abdus Salam Centre For Theoretical Physics, Trieste Italy for support and hospitality throughout these years. The author is also grateful to the referee who examined the previous version thoroughly and offered numerous invaluable suggestions.


\newpage

\bibliographystyle{phaip}

\end{document}